\documentclass[twocolumn,twocolappendix,times]{aastex631}

\usepackage{CJK}
\usepackage{appendix}

\newcommand{\se}{\textsc{source extractor}}
\newcommand{\sep}{\textsc{sep}}
\newcommand{\tphot}{\textsc{tphot}}
\newcommand{\pipeline}{\textsc{pipeline}}

\newcommand{\astrorms}{\textsc{astrorms}}
\newcommand{\python}{\textsc{python}}

\newcommand{\fst}[1]{#1}

\shorttitle{CEERS MIRI Imaging}
\shortauthors{Yang et al.}

\graphicspath{{./}{figures/}}

\begin{document}
\begin{CJK*}{UTF8}{gbsn}

\title{CEERS MIRI Imaging: Data Reduction and Quality Assessment}

\correspondingauthor{Guang Yang}
\email{gyang206265@gmail.com}

\author[0000-0001-8835-7722]{Yang, G. (杨光)}
\affiliation{Kapteyn Astronomical Institute, University of Groningen, P.O. Box 800, 9700 AV Groningen, The Netherlands}
\affiliation{SRON Netherlands Institute for Space Research, Postbus 800, 9700 AV Groningen, The Netherlands}

\author[0000-0001-7503-8482]{Papovich, C.}
\affiliation{Department of Physics and Astronomy, Texas A\&M University, College
Station, TX, 77843-4242 USA}
\affiliation{George P.\ and Cynthia Woods Mitchell Institute for
 Fundamental Physics and Astronomy, Texas A\&M University, College Station, TX, 77843-4242 USA}

\author[0000-0002-9921-9218]{Bagley, M. B.}
\affiliation{Department of Astronomy, The University of Texas at Austin, Austin, TX, USA}

\author[0000-0001-7113-2738]{Ferguson, H. C.}
\affiliation{Space Telescope Science Institute, Baltimore, MD, USA}

\author[0000-0001-8519-1130]{Finkelstein, S. L.}
\affiliation{Department of Astronomy, The University of Texas at Austin, Austin, TX, USA}

\author[0000-0002-6610-2048]{Koekemoer, A. M.}
\affiliation{Space Telescope Science Institute, 3700 San Martin Dr., Baltimore, MD 21218, USA}

\author[0000-0003-4528-5639]{P\'erez-Gonz\'alez, P. G.}
\affiliation{Centro de Astrobiolog\'{\i}a (CAB), CSIC-INTA, Ctra. de Ajalvir km 4, Torrej\'on de Ardoz, E-28850, Madrid, Spain}

\author[0000-0002-7959-8783]{Arrabal Haro, P.}
\affiliation{NSF's National Optical-Infrared Astronomy Research Laboratory, 950 N. Cherry Ave., Tucson, AZ 85719, USA}

\author[0000-0003-0492-4924]{Bisigello, L.}
\affiliation{Dipartimento di Fisica e Astronomia "G.Galilei", Universit\'a di Padova, Via Marzolo 8, I-35131 Padova, Italy}
\affiliation{INAF--Osservatorio Astronomico di Padova, Vicolo dell'Osservatorio 5, I-35122, Padova, Italy}

\author[0000-0001-8183-1460]{Caputi, K. I.}
\affiliation{Kapteyn Astronomical Institute, University of Groningen, P.O. Box 800, 9700 AV Groningen, The Netherlands}

\author[0000-0001-8551-071X]{Cheng, Y.}
\affiliation{University of Massachusetts Amherst, 710 North Pleasant Street, Amherst, MA 01003-9305, USA}

\author[0000-0001-6820-0015]{Costantin, L.}
\affiliation{Centro de Astrobiolog\'ia (CAB), CSIC-INTA, Ctra de Ajalvir km 4, Torrej\'on de Ardoz, 28850, Madrid, Spain}

\author[0000-0001-5414-5131]{Dickinson, M.}
\affiliation{NSF's National Optical-Infrared Astronomy Research Laboratory, 950 N. Cherry Ave., Tucson, AZ 85719, USA}

\author[0000-0003-3820-2823]{Fontana, A.}
\affiliation{INAF - Osservatorio Astronomico di Roma, via di Frascati 33, 00078 Monte Porzio Catone, Italy}

\author[0000-0003-2098-9568]{Gardner, J. P.}
\affiliation{Astrophysics Science Division, Goddard Space Flight Center, Code 665, Greenbelt, MD 20771, USA}

\author[0000-0002-5688-0663]{Grazian, A.}
\affiliation{INAF--Osservatorio Astronomico di Padova, Vicolo dell'Osservatorio 5, I-35122, Padova, Italy}

\author[0000-0001-9440-8872]{Grogin, N. A.}
\affiliation{Space Telescope Science Institute, 3700 San Martin Dr., Baltimore, MD 21218, USA}

\author[0000-0003-0129-2079]{Harish, S.}
\affiliation{Laboratory for Multiwavelength Astrophysics, School of Physics and Astronomy, Rochester Institute of Technology, 84 Lomb Memorial Drive, Rochester, NY 14623, USA}

\author[0000-0002-4884-6756]{Holwerda, B. W.}
\affil{Physics \& Astronomy Department, University of Louisville, 40292 KY, Louisville, USA}

\author[0000-0001-8386-3546]{Iani, E.}
\affiliation{Kapteyn Astronomical Institute, University of Groningen, P.O. Box 800, 9700 AV Groningen, The Netherlands}

\author[0000-0001-9187-3605]{Kartaltepe, J. S.}
\affiliation{Laboratory for Multiwavelength Astrophysics, School of Physics and Astronomy, Rochester Institute of Technology, 84 Lomb Memorial Drive, Rochester, NY 14623, USA}

\author[0000-0001-8152-3943]{Kewley, L. J.}
\affiliation{Center for Astrophysics | Harvard \& Smithsonian, 60 Garden Street, Cambridge, MA 02138, USA}

\author[0000-0002-5537-8110]{Kirkpatrick, A.}
\affiliation{Department of Physics and Astronomy, University of Kansas, Lawrence, KS 66045, USA}

\author[0000-0002-8360-3880]{Kocevski, D. D.}
\affiliation{Department of Physics and Astronomy, Colby College, Waterville, ME 04901, USA}

\author[0000-0002-5588-9156]{Kokorev, V.}
\affiliation{Kapteyn Astronomical Institute, University of Groningen, P.O. Box 800, 9700 AV Groningen, The Netherlands}

\author[0000-0003-3130-5643]{Lotz, J. M.}
\affiliation{Gemini Observatory/NSF's National Optical-Infrared Astronomy Research Laboratory, 950 N. Cherry Ave., Tucson, AZ 85719, USA}

\author[0000-0003-1581-7825]{Lucas, R. A.}
\affiliation{Space Telescope Science Institute, 3700 San Martin Drive, Baltimore, MD 21218, USA}

\author[0000-0001-6066-4624]{Navarro-Carrera, R.}
\affiliation{Kapteyn Astronomical Institute, University of Groningen, P.O. Box 800, 9700 AV Groningen, The Netherlands}

\author[0000-0001-8940-6768]{Pentericci, L.}
\affiliation{INAF - Osservatorio Astronomico di Roma, via di Frascati 33, 00078 Monte Porzio Catone, Italy}

\author[0000-0003-3382-5941]{Pirzkal, N}
\affiliation{ESA/AURA Space Telescope Science Institute}

\author[0000-0002-5269-6527]{Ravindranath, S.}
\affiliation{Space Telescope Science Institute, 3700 San Martin Drive, Baltimore, MD 21218, USA}

\author{Rinaldi, P.}
\affiliation{Kapteyn Astronomical Institute, University of Groningen, P.O. Box 800, 9700 AV Groningen, The Netherlands}

\author[0000-0001-9495-7759]{Shen, L.}
\affiliation{Department of Physics and Astronomy, Texas A\&M University, College
Station, TX, 77843-4242 USA}
\affiliation{George P.\ and Cynthia Woods Mitchell Institute for
 Fundamental Physics and Astronomy, Texas A\&M University, College
 Station, TX, 77843-4242 USA}

\author[0000-0002-6748-6821]{Somerville, R. S.}
\affiliation{Center for Computational Astrophysics, Flatiron Institute, 162 5th Avenue, New York, NY, 10010, USA}

\author[0000-0002-1410-0470]{Trump, J. R.}
\affiliation{Department of Physics, 196 Auditorium Road, Unit 3046, University of Connecticut, Storrs, CT 06269, USA}

\author[0000-0002-6219-5558]{de la Vega, A.}
\affiliation{Department of Physics and Astronomy, University of California, 900 University Ave, Riverside, CA 92521, USA}

\author[0000-0003-3903-6935]{Wilkins, S. M.}
\affiliation{Astronomy Centre, University of Sussex, Falmer, Brighton BN1 9QH, UK}
\affiliation{Institute of Space Sciences and Astronomy, University of Malta, Msida MSD 2080, Malta}

\author[0000-0003-3466-035X]{Yung, L. Y. A.}
\affiliation{Astrophysics Science Division, NASA Goddard Space Flight Center, 8800 Greenbelt Rd, Greenbelt, MD 20771, USA}





\begin{abstract}
The Cosmic Evolution Early Release Science Survey (CEERS), targeting the Extended Groth Strip extragalactic field, is one of the \textit{JWST} Director's Discretionary Early Release Science programs.
To date, all observations have been executed and include NIRCam/MIRI imaging and NIRSpec/NIRCam spectroscopic exposures. 
Here, we discuss the MIRI imaging, which includes eight pointings, four of which provide deep imaging with the bluer bands (F560W, F770W) and four with contiguous wavelength coverage in F1000W, F1280W, F1500W, and F1800W, where two of these   also include coverage in F770W and F2100W.  
We present a summary of the data, the data quality and data reduction.  The data reduction is based on the \textsc{jwst calibration pipeline} combined with custom modifications and additional steps designed to enhance the output quality, including improvements in astrometry and the removal of detector artifacts.
We estimate the image depth of the reduced mosaics, and show that these generally agree with expectations from the Exposure Time Calculator.
%
%
We compare the MIRI F560W and F770W flux densities for bright sources to measurements from \textit{Spitzer}/IRAC Ch3 (5.8~$\mu$m) and Ch4 (8.0~$\mu$m), and we find that they agree with systematic differences of $<0.1$~mag.
For the redder MIRI bands, we assess their quality by studying the spectral energy distributions (SEDs) of Galactic stars. 
The SEDs are consistent with the expected Rayleigh--Jeans law with a deviation $\sim 0.03$~mag, indicating that the MIRI colors are reliable. 
We also discuss all publicly released data products (images and source catalogs), which are available on the CEERS website (\url{https://ceers.github.io/}).

\end{abstract}

\section{Introduction} \label{sec:intro}

Mid-infrared (MIR) observations provide critical insights for extragalactic astronomy.
Past measurements of key extragalactic datasets have established that at least half of the light from star formation and accretion onto supermassive black holes (SMBHs) is absorbed by molecules and carbon-/silicon-based ``dust'', and re--radiated at longer wavelengths ($\sim$8--1000~\micron; see reviews by, e.g., \citealt{madau14,hickox18}).  
Star formation, when obscured by dust, has weak or non-detectable emission in the UV/optical. 
However, obscured star formation can often be revealed by the strong polycyclic aromatic hydrocarbon (PAH) emission features at the rest-frame wavelengths of $\sim 3$--18~$\mu$m \citep[e.g.,][]{spoon07, tielens08, shipley16}.
Similarly, it is also infeasible to identify heavily obscured type~2 active galactic nuclei (AGN) in the UV/optical. 
If the absorption column density ($N_{\rm H}$) reaches the Compton-thick regime ($\gtrsim 10^{24}$~cm$^{-2}$), even hard \hbox{X-rays} cannot effectively penetrate through the large amounts of obscuring materials \citep[e.g.,][]{brandt15, brandt22}.
Instead, significant MIR emission (peaking at $\sim 10$~$\mu$m) is a prevalent feature among AGN (whether obscured or not), and thus MIR selections are efficient to identify the hidden AGN population missed in other wavelengths \citep[e.g.,][]{kirkpatrick12, kirkpatrick15, hickox18}.

Motivated by the importance of MIR observations, many space-borne MIR missions, such as \textit{Infrared Astronomical Satellite (IRAS)}, \textit{Infrared Space Observatory (ISO)},  \textit{Spitzer Space Telescope (Spitzer)}, \textit{AKARI}, and \textit{Wide-field Infrared Survey Explorer (WISE)}, were successfully deployed in the past four decades. 
These missions have significantly advanced our understanding of extragalactic astronomy. 
For example, \textit{Spitzer} played a crucial role in selecting the highest redshift galaxies and constraining their properties (e.g., GN-z11; \citealt{oesch16}).
\textit{WISE}\ discovered a new type of MIR hyper-luminous object, hot dust-obscured galaxies, representing a stage of powerful SMBH accretion and/or star formation at $z\gtrsim 2$ \citep[e.g.,][]{eisenhardt12, tsai15}. 

The Mid-Infrared Instrument (MIRI; \citealt{wright23}) onboard the recently deployed \textit{James Webb Space Telescope} (\textit{JWST}; \citealt{gardner23}) has unprecedented sensitivity and angular resolution at the MIR wavelengths of $\sim 5$--28~$\mu$m \citep{rieke15, glasse15}.
MIRI is a versatile instrument with the capabilities of broad-band imaging \citep{bouchet15}, coronagraphy \citep{boccaletti15}, slit/slitless low-resolution spectroscopy \citep[LRS;][]{kendrew15}, and integral-field medium-resolution spectroscopy \citep[MRS;][]{wells15}.
MIRI is equipped with 3 arsenic-doped silicon impurity band conduction (Si:As IBC) detectors, all in a $1024\times 1032$ format \citep{rieke15}.
Two detectors are used exclusively by the Integral field spectroscopy (IFS), and one detector is shared by the other observational modes. 
Detailed simulations show that MIRI is able to probe into the physical nature of the faintest star-forming galaxies and AGN in the distant Universe \citep[e.g.,][]{bisigello17, kirkpatrick17, satyapal21, yang21}.
After the successful deployment of \textit{JWST}, MIRI has already started to show its excellent power to study the MIR Universe (e.g., \citealt{iani22, akins23, alvarez23, colina23, kirkpatrick23, magnelli23, rinaldi23, papovich23, shen23, yang23}).

The Cosmic Evolution Early Release Science (CEERS, proposal \#1345, PI: S. Finkelstein) is a \textit{JWST}\ extragalactic survey program.
CEERS consists of $\sim 68$ hours of NIRCam/MIRI imaging and NIRSpec/NIRCam spectroscopic observations, covering a total area of $\sim 100$~arcmin$^2$ in the Extended Groth Strip (EGS) field. 
The main science goals of CEERS include (but not limited to) probing the first galaxies and AGN, tracing the galaxy/SMBH assembly across the cosmic history, and establishing the near-infrared (NIR) emission-line and MIR continuum diagnostics of distant galaxies/AGN.  
All planned CEERS observations were performed successfully, and have been used to demonstrate the ability of \textit{JWST} to study the properties of distant galaxies \citep[e.g.,][]{finkelstein22, finkelstein23, kocevski23, larson23, papovich23, zavala23}. 

In this paper, we discuss the MIRI imaging component of CEERS.  
We describe the observations and the properties of the MIRI datasets. We also discuss the details of the data reduction, the properties of the final images, and source catalogs, which we provide with this publication.  
The structure of this paper is as follows.
In \S\ref{sec:analysis}, we describe the observations and our data reduction methods.  
In \S\ref{sec:assess}, we assess the quality of our reduced data products.  
We summarize our results and discuss future prospects in \S\ref{sec:sum}.

Throughout this paper, all magnitudes are in AB units \citep{oke83}, where $m_\mathrm{AB} = -48.6 - 2.5 \log(f_\nu)$ for $f_\nu$ in units of erg~s$^{-1}$~cm$^{-2}$ Hz$^{-1}$.  
Quoted uncertainties are at the $1\sigma$\ (68\%) confidence level.

\section{Observations and data reduction}
\label{sec:analysis}

\subsection{Observations}
\label{sec:obs}
There are eight MIRI pointings in CEERS, named MIRI1, MIRI2, MIRI3, MIRI6, MIRI5, MIRI7, MIRI8, and MIRI9.\footnote{The numbering scheme is based on all CEERS pointings, where these eight pointings have MIRI components. Note that the numbering scheme has been updated to reflect the final CEERS observational configuration. For example, MIRI4 was in an older version of the CEERS observation plan but replaced by MIRI9 later when the observations were split into two epochs.}
Table~\ref{tab:obs} summarizes the basic properties for each pointing.

These MIRI observations were performed in two epochs. 
Epoch~1 and Epoch~2 occurred in June 2022 and December 2022, respectively.  
MIRI1, MIRI2, MIRI3, and MIRI6 were observed in Epoch~1 as prime observations with parallel NIRCam imaging. 
The NIRCam imaging observations and data reduction are described in detail by \cite{bagley22}.  
MIRI5, MIRI7, MIRI8, and MIRI9 were observed in Epoch~2, in parallel while NIRCam Wide Field Slitless Spectroscopy (WFSS) was prime.

The dithers of Epoch~1 and Epoch~2 observations are standard 3-point patterns for MIRI/Imaging $+$ NIRCam/Imaging and 4-point patterns of NIRCam/WFSS $+$ MIRI/Imaging, respectively.\footnote{\url{https://jwst-docs.stsci.edu/jppom/parallel-observations/coordinated-parallel-observations/additional-dithers-for-coordinated-parallel-observations}}
The Epoch~1 exposures are relatively deeper compared to the Epoch~2 ones in general due to the fact that we structured the MIRI observations to match the CEERS NIRCam imaging and WFSS integration times, which results in different amounts of available time for the observations (see Table~\ref{tab:obs}).

In terms of wavelength coverage, MIRI1, MIRI2, MIRI5, and MIRI8 are ``red'' pointings having F1000W, F1280W, F1500W, and F1800W. 
MIRI1 and MIRI2 also have coverage with F770W and F2100W. 
These red MIRI pointings all overlap with the \textit{Hubble Space Telescope (HST)} CANDELS imaging \citep[][]{grogin11,koekemoer11} and catalogs \citep[e.g.,][]{stefanon17,barro19}, with less (or no) overlap with the CEERS/NIRCam imaging.
These fields thereby have contiguous wavelength coverage with MIRI imaging from at least 10 to 18~$\mu$m (with MIRI 1 and MIRI 2 also having coverage extending to 7.7 and 21~$\mu$m).   
The principal science drivers for these pointings are to understand the MIR spectral energy distributions (SEDs) of $0.5 \lesssim z \lesssim 3$ galaxies detected in the \textit{HST}\ images \citep[e.g.,][]{shen23,yang23}. 

The other pointings, MIRI3, MIRI6, MIRI7, and MIRI9 are ``blue'' pointings having F560W and F770W.  
These pointings are generally deeper (in terms of AB mag) and overlap with the CEERS/NIRCam pointings.  
The principal science driver here was to understand the rest-frame optical/NIR SEDs of galaxies at $4 \lesssim z \lesssim 10$ \citep[e.g.,][]{papovich23,barro23}.  

When constructing the observations with the Astronomer's Proposal Tool (APT), we encountered a ``data-rate overflow'' warning for the F560W and F770W exposures in the typical ``FASTR1'' readout mode. 
To avoid this warning, we changed to the ``SLOWR1'' readout mode which effectively reduced the data rate. 
Therefore, the F560W and F770W exposures were performed with SLOWR1, while the exposures of the other MIRI bands were performed with FASTR1 mode.

\begin{table*}
\centering
\caption{CEERS MIRI Observations}
\label{tab:obs}
\begin{tabular}{ccccc} \hline\hline
Field & Center Position & Date & Parallel & Filters (Exposure time) \\
\hline 
MIRI1 & 14:20:38.9, $+$53:03:04.6 & June 2022 & NIRCam/Imaging & F770W (1648 s), F1000W (1673 s), F1280W (1673 s), \\ & & & & F1500W (1673 s), F1800W (1698 s), F2100W (4812 s) \\ 
MIRI2 & 14:20:17.4, $+$52:59:16.2 & June 2022 & NIRCam/Imaging & F770W (1648 s), F1000W (1673 s), F1280W (1673 s), \\ & & & & F1500W (1673 s), F1800W (1698 s), F2100W (7884 s) \\
MIRI3 & 14:19:55.2, $+$52:55:29.4 & June 2022 & NIRCam/Imaging & F560W (2938 s), F770W (8815 s) \\
MIRI6 & 14:20:07.8, $+$52:54:21.8 & June 2022 & NIRCam/Imaging & F560W (2938 s), F770W (8815 s) \\
MIRI5 & 14:19:05.2, $+$52:49:27.5 & Dec. 2022 & NIRCam/WFSS    & F1000W (1243 s), F1280W (932 s), F1500W (932 s), F1800W (1243 s) \\
MIRI8 & 14:19:22.6, $+$52:49:14.1 & Dec. 2022 & NIRCam/WFSS    & F1000W (1243 s), F1280W (932 s), F1500W (932 s), F1800W (1243 s) \\
MIRI7 & 14:19:45.1, $+$52:53:03.8 & Dec. 2022 & NIRCam/WFSS    & F560W (1433 s), F770W (2580 s) \\
MIRI9 & 14:19:00.2, $+$52:45:28.0 & Dec. 2022 & NIRCam/WFSS    & F560W (1433 s), F770W (2580 s) \\
\hline
\end{tabular}
\end{table*}

\subsection{Data reduction}
\label{sec:reduction}
\begin{figure*}[tp]
    \centering
    \includegraphics[width=1\textwidth]{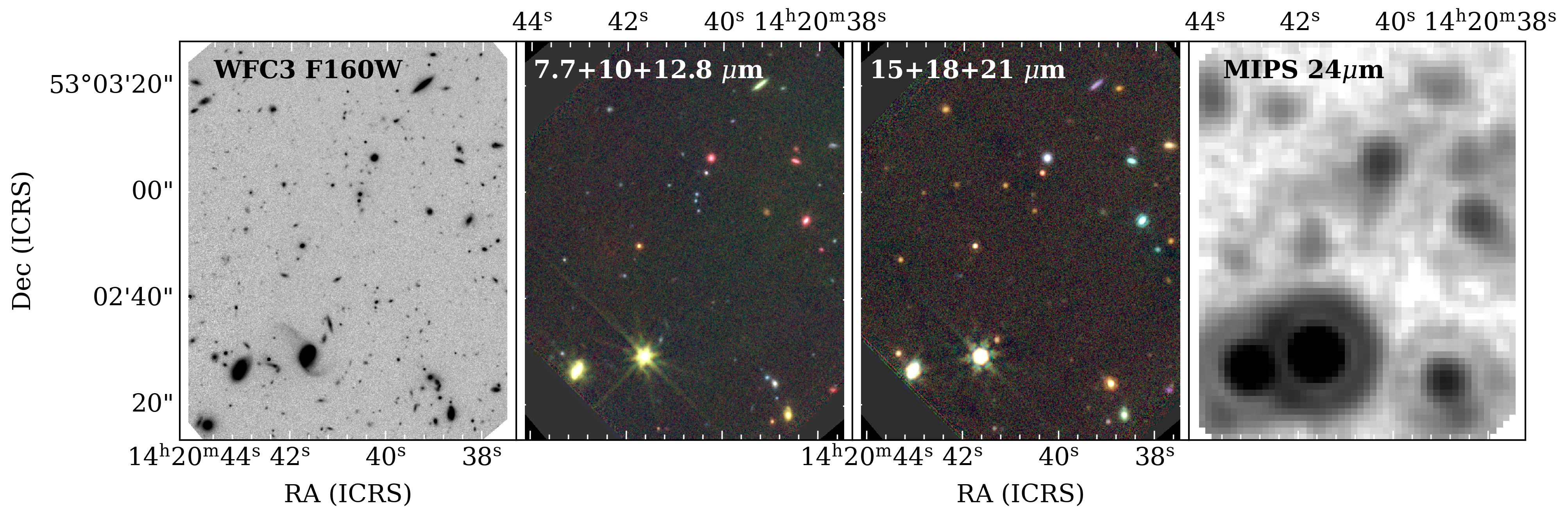}
    \caption{An illustration of the MIRI imaging in field MIRI1 based on the reduced data products from this work compared to the existing CANDELS WFC3 F160W imaging and \textit{Spitzer}/MIPS 24~$\mu$m imaging.  From left to right the panels show a portion of the imaging from: \textit{HST} F160W;  MIRI F770W (blue) $+$ F1000W (green) $+$ F1280W (red); MIRI F1500W (blue) $+$ F1800W (green) $+$ F2100W (red); and MIPS 24~$\mu$m. 
    The depth, image-quality, and contiguous wavelength coverage from CEERS/MIRI allow us to see a range of mid-IR ``colors'' indicative of mid-IR emission features  redshift through the different MIRI bands.
    }
    \label{fig:rgb}
\end{figure*}

We first downloaded the raw MIRI data (``uncal.fits'' files) from MAST.\footnote{\url{https://mast.stsci.edu/portal/Mashup/Clients/Mast/Portal.html}}
We then processed the raw data using the {\sc jwst calibration pipeline} v1.10.2 (\pipeline\ hereafter; \fst{\citealt{bushouse22}}) with \textit{JWST}\ Calibration Reference Data System (CRDS) context jwst\_1077.pmap.  The
\pipeline\ has three stages of reduction.
Below, we detail our reduction procedures for these three stages, which primarily follow the default pipeline, but we add some additional custom steps.  
These are summarized in Table~\ref{tab:reduction}. 
\fst{Our reduction code is publicly available at \url{https://github.com/ceers/ceers-miri} \citep{my_miri_code_doi}.}
The reduction yields science-ready images and associated root-mean-square (RMS) maps for all available fields/bands.
Fig.~\ref{fig:rgb} displays MIRI RGB images based on the reduced mosaics, in comparison with the \textit{HST}/WFC3 F160W catalog\footnote{\url{http://doi.org/10.17909/T94S3X}} \citep{koekemoer11} and \textit{Spitzer}/MIPS 24~$\mu$m for one of the CEERS red MIRI pointings (MIRI1).

\begin{table*}
\centering
\caption{CEERS MIRI Reduction}
\label{tab:reduction}
\begin{tabular}{lll} 
\hline
\hline
Procedure & Functionality & Section \\
\hline 
Stage 1 & Perform ramp fits and produce count-rate maps & \ref{sec:stage1} \\
Jump masking$^a$ & Identify and mask corrupted jump groups in F2100W & \ref{sec:stage1} \\
Stage 2 & Calibrate count-rate maps and output a science image for each dither & \ref{sec:stage2} \\
Stripe removal & Remove the horizontal and vertical stripes & \ref{sec:stage2} \\
Super background & Subtract background using the super-background algorithm & \ref{sec:stage2} \\
Latent removal$^b$ & Mask the pixels affected by latent artifacts & \ref{sec:stage2} \\
TweakReg & Correct the astrometry based on the \textit{HST} reference & \ref{sec:stage3} \\
Outlier Detection & Reject remaining cosmic rays & \ref{sec:stage3} \\
Resampling & Combine dithers into a final mosaic aligned with the \textit{HST} WCS frame & \ref{sec:stage3} \\
RMS map & Produce properly scaled RMS maps using \textsc{astrorms} & \ref{sec:stage3} \\
\hline 
\end{tabular}
\begin{flushleft}
{\sc Note.} --- 
(a) This is applied before the ramp-fit step in Stage~1. 
(b) This is only applied to the June~28th (2022) exposure of MIRI2/F2100W.
\fst{Our custom reduction code is publicly available at \url{https://github.com/ceers/ceers-miri}.}
\end{flushleft}
\end{table*}

\subsubsection{Stage 1}
\label{sec:stage1}
Stage~1 performs detector-level processing that reduces each four-dimensional (x-y-group-integration) ramp data to a two-dimensional (x-y) rate map. 
Based on our experience with simulated MIRI data \citep{yang21}, the default stage~1 parameters are in general sufficient and do not require adjustments.
Therefore, we ran the stage~1 \pipeline\ with the default parameters for all exposures except for two segments of F2100W in MIRI1 and MIRI2 (for reasons discussed in the next paragraph).

\begin{figure*}[htp]
    \centering
	\includegraphics[width=2\columnwidth]{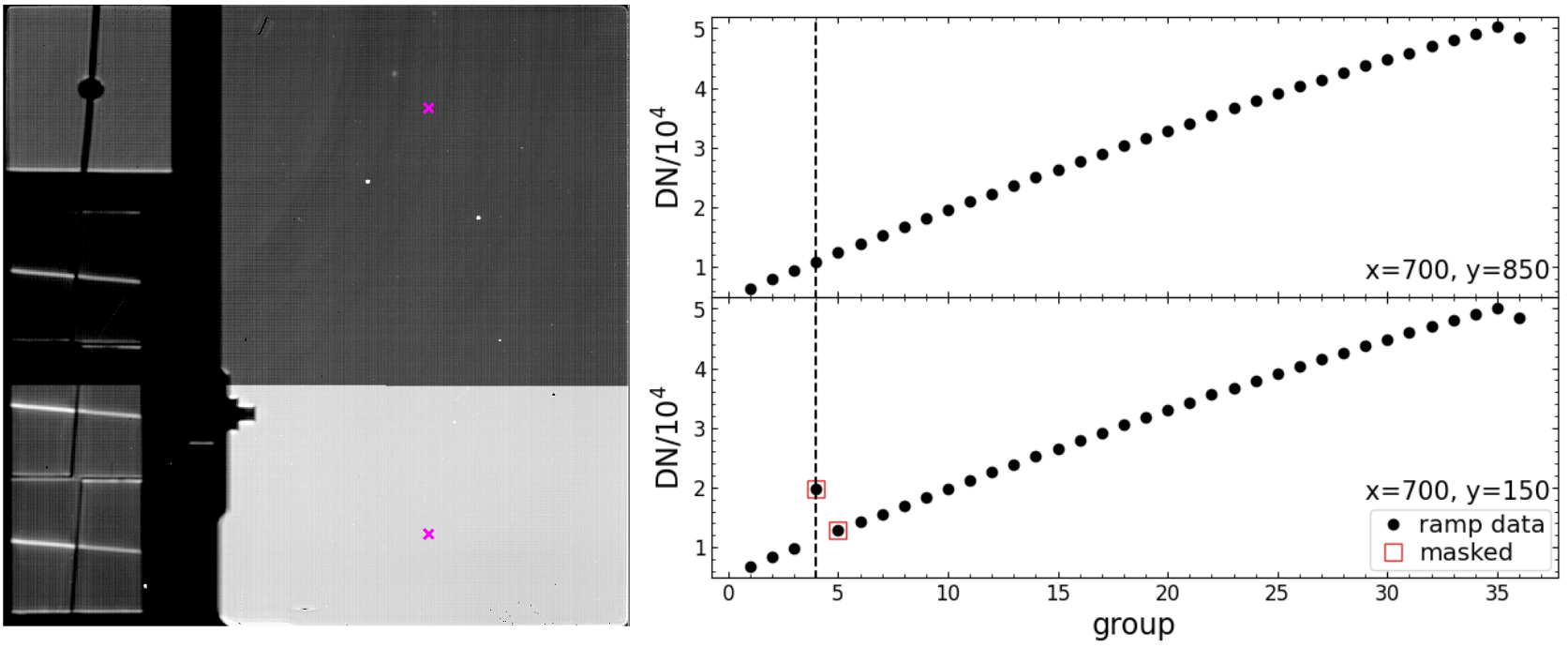}
    \caption{Example of corrupted jumps in one of our F2100W dithers.
    The left panel displays a group in the raw ramp data, where the data issue happens on the bottom rows of the detector.
    The right panels display the ramp data (digital number, DN, versus group) in all of the groups for the two pixels (x-y positions labeled) marked as magenta ``x'' on the left. 
    The dashed vertical line indicates the group displayed on the left.
    The red squares indicate the data affected by the data issue, and all such data are identified and masked out in our reduction.  
    Note that the dips in the last group (group 36 in this example) are discarded by default in stage~1 \pipeline.
    }
    \label{fig:jump_2pix}
\end{figure*}

During the observations for MIRI 1 and MIRI 2, segments of the F2100W data  experienced issues and were partially corrupted (STScI, J.\ Morrison, private communication).  
This impacted 3072~s (out of 4812~s) of the data in MIRI1/F2100W and 3072~s (out of 7884~s) in MIRI2/F2100W. 
Specifically, the issue causes some corrupted ramp data, manifesting as ``jumps'' in the ramp (``corrupted jumps'', hereafter). 
Compared to the cosmic-ray jumps, which affect some random pixels, these corrupted jumps typically happen simultaneously across $\gtrsim 100$ consecutive rows or even the entire detector. 
Fig.~\ref{fig:jump_2pix} displays an example of the corrupted jumps. 
Each jump affects $\sim 2$--3 (out of 36) groups in one integration. 
About half of the integrations in the two segments of F2100W data have this issue. 

The stage~1 \pipeline\ has a ``Jump'' step to mask out jumps caused by cosmic rays \fst{\citep{morrison23}}. 
However, this step is not able to detect all the corrupted jumps in the datasets impacted by this issue.
Therefore, we wrote a custom \python\ code to detect and mask the corrupted jumps. 
This code searches for the corrupted jumps based on their multi-row feature.
From our visual inspection, it successfully detects all corrupted jumps in the F2100W data. 
We apply our code right after the ``Jump'' step and before the ``Ramp Fitting'' step in this stage of the pipeline. 
As a sanity check, we furthermore tested our code on other ``good'' data (unaffected by this issue) and do not detect any corrupted jumps. 

\subsubsection{Stage 2}
\label{sec:stage2}
Stage~2 reduces each rate map from stage~1 to a calibrated science image. 
We ran the stage-2 \pipeline\ with the default parameters for F1000W, F1280W, F1500W, F1800W, and F2100W. 
For F560W and F770W, the data were obtained in SLOWR1 readout mode (\S\ref{sec:obs}).
However, at the time when we carried out this work, the SLOWR1 flat-field CRDS files for these two bands were based on ground tests rather on inflight calibrations, and thus they were severely outdated.
The FASTR1 CRDS files are considered better than these SLOWR1 ones (\fst{Karl Gordon and \textit{JWST} helpdesk, private communications}).
Therefore, we changed the default SLOWR1 flat-field CRDS files to the corresponding FASTR1 files when running stage~2 for F560W and F770W.\footnote{\fst{Note that this change only applies to flat fielding. 
For other processing steps such as dark-current subtraction, we used the default SLOWR1 CRDS files which are up-to-date.}}
This is a temporary solution, and in the future we intend to adopt the in--flight SLOWR1 files when they become available. 
We expect the systematical changes after applying the in--flight SLOWR1 files in the future will be minor ($\lesssim 0.1$~mag), as the MIRI photometry from the current reduction already agrees well with \textit{Spitzer/IRAC} (see \S\ref{sec:spitzer}).

\begin{figure}[htp]
    \centering
	\includegraphics[width=\columnwidth]{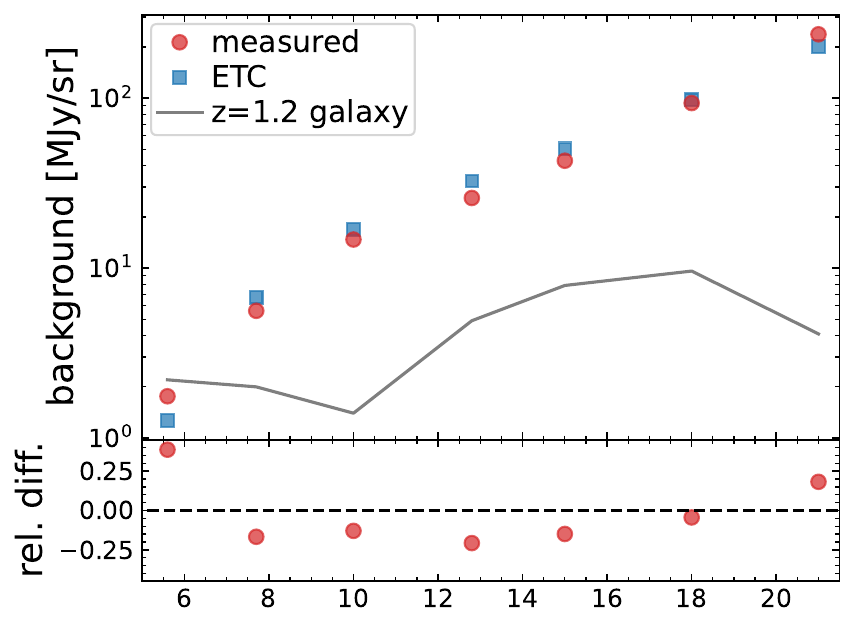}
    \caption{\textit{Top}: Measured background (red points) versus ETC predicted values (blue squares). 
    The measured background is the median value of all dithers/exposures. 
    Both backgrounds rise strongly toward longer wavelengths.
    The measured background is similar to the ETC prediction except at F2100W, where the former is significantly higher.
    This offset could lead to that our measured F2100W depth is shallower than the ETC depth (\S\ref{sec:depth}).
    For reference, the grey line represents the surface brightness (as measured by the central MIRI pixel) of a simulated massive ($\log M_\ast/M_\odot = {10.6}$\ ) star-forming (SFR = 15 $M_\odot$~yr$^{-1}$) galaxy at $z=1.2$ \citep{yang21}.
    \textit{Bottom}: the relative difference between the measured and ETC backgrounds, i.e., (measured $-$ ETC) / ETC.  
    }
    \label{fig:bkg}
\end{figure}

We estimate the background in units of MJy/sr by calculating the median pixel values of the stage~2 output images.
The background variation among different dithers/exposures is typically within a few percent. 
We compare the median background values of all dithers/exposures versus the Exposure Time Calculator (ETC) predictions in Fig.~\ref{fig:bkg}.
The ETC predictions were calculated assuming the same sky coordinates and observation dates as the real exposures.      
For F1000W, F1280W, F1500W, and F1800W, the measured background is slightly lower than the ETC prediction by 0-20\%.
In contrast, for F560W and F2100W, the measured background is higher than the ETC prediction by 39\% and 18\%, respectively. 
For reference, Fig.~\ref{fig:bkg} also displays a grey line representing the surface brightness (as measured by the central MIRI pixel) of a simulated massive ($\log M_\ast/M_\odot = 10.6$) star-forming (SFR = 15\ $M_\odot$~yr$^{-1}$) galaxy at $z=1.2$ \citep{yang21}.
The background emission dominates the source emission in all MIRI bands except F560W, because the former rises strongly toward longer wavelengths.

\begin{figure*}[htp]
    \centering
	\includegraphics[width=2\columnwidth]{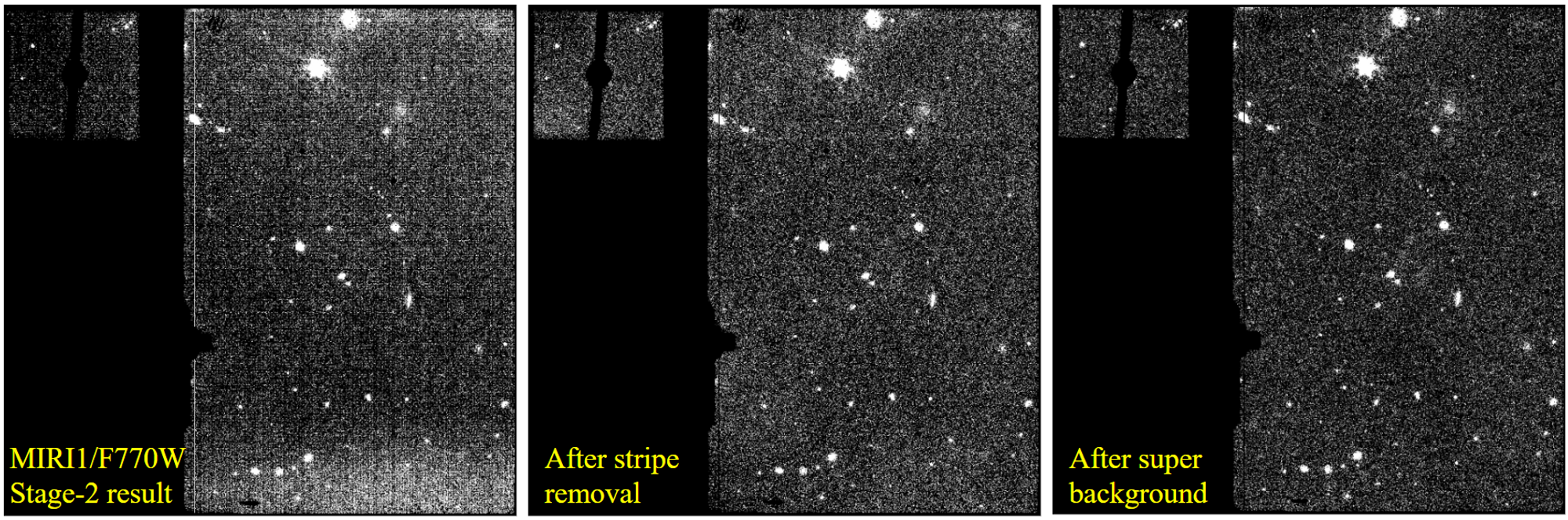}
    \caption{A MIRI1/F770W dither image from the default Stage~2 pipeline (left), after our stripe removal (middle), and after our super-background noise subtraction described in \S\ref{sec:stage2}.
    The three images are on the same color scale. 
    The noise becomes less prominent after our procedures of stripe removal and super background. 
    }
    \label{fig:noise}
\end{figure*}

Upon viewing the images produced by stage~2, we noted some obvious ``stripe-like'' noise patterns in the vertical and/or horizontal directions (e.g., Fig.~\ref{fig:noise} left; \fst{\citealt{morrison23}}). 
To remove such noise, we first subtracted each detector row with the median value of its pixels, adopting a sigma clip (2$\sigma$) to avoid pixels illuminated by sources. 
We did a similar procedure for each column. 
After this step, the stripes are largely removed, but there are still some residual noise patterns (e.g., Fig.~\ref{fig:noise} center).
The noise patterns are similar in all dithers, in terms of detector coordinates (x,y).
To address such noise, we implemented a ``super-background'' algorithm (also described in \citealt{papovich23}).
We estimated a ``super-background'' frame for each dither. 
A pixel in this frame is calculated as the median pixel value of all other dithers of the same band (including dithers in all pointings of the same epoch). 
In this step, we also masked out bad pixels and those affected by sources using \sep\ (version 1.2.1; \citealt{barbary16}).
Finally, we subtracted the super-background frame from the science image. 
Fig.~\ref{fig:noise} (right) displays the image after our super-background subtraction, where the noise patterns are largely removed. 

Another unusual thing we find is a bright point source in the three MIRI2/F2100W dithers obtained on June 28th, but this source does not exist in the other MIRI2 (F2100W or other bands) exposures.
The centroid of this source has a fixed detector (x,y) coordinate and a varying physical (RA,DEC) coordinate in the three dithers. 
Also, its flux in the three dithers declines chronologically.
Therefore, we consider this source as a detector artifact, i.e., latent (persistence), which is common among IR detectors \fst{\citep[e.g.,][]{rieke07, morrison23}}.
By carefully inspecting the three dithers, we find additional fainter sources in these observations with similar behaviors.
We perform a MAST search but do not find a ``previous'' MIRI observation that is responsible for the latent artifacts in our data, and thus the latent was likely produced when MIRI was not in use (neither as in primary nor in parallel) or during a slew process, but was still collecting photons.

To address the issue of latent sources, we mask the affected pixels following the procedures below. 
We first detect sources in each dither using \sep.  
Flux associated with latent sources do not move in x-y coordinate (so remain fixed on the detector compared to real sources, which move with the dither pattern). 
The \sep\ run leads to three source lists and segmentation maps corresponding to the three dithers. 
For each object detected in the 1st dither, we search for its counterpart within a radius of $0.5''$ in the other 2 dithers using the physical coordinates (RA,DEC).
If any counterpart is found, we consider this object as a real physical source; otherwise, we consider it as an artifact of a latent source.
We produce a mask map by removing the real sources in the first dither's segmentation map. 
We use the first dither, because it should have the strongest latent effects.
The mask map is applied to all three dithers to clean up the pixels (potentially) affected by latent artifacts.

\subsubsection{Stage 3}
\label{sec:stage3}
Stage~3 performs additional corrections for each dither image produced by Stage 2, and combine different dithers into a final multi-extension FITS file including, e.g., a science image and a weight map.

\begin{figure*}[htp]
    \centering
	\includegraphics[width=\columnwidth]{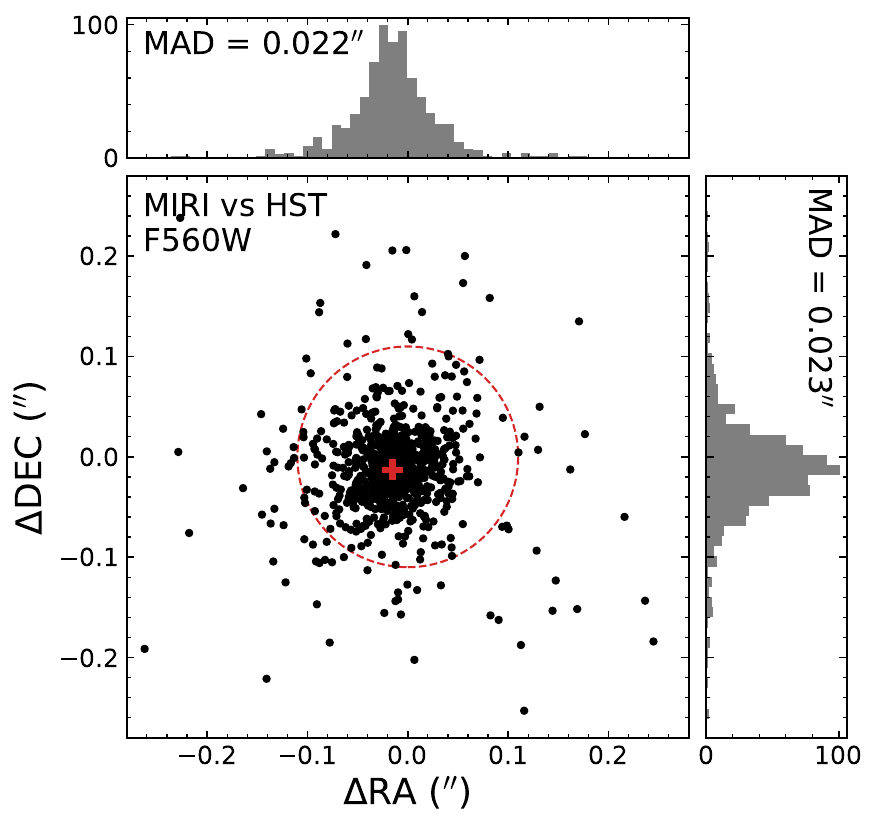}
	\includegraphics[width=\columnwidth]{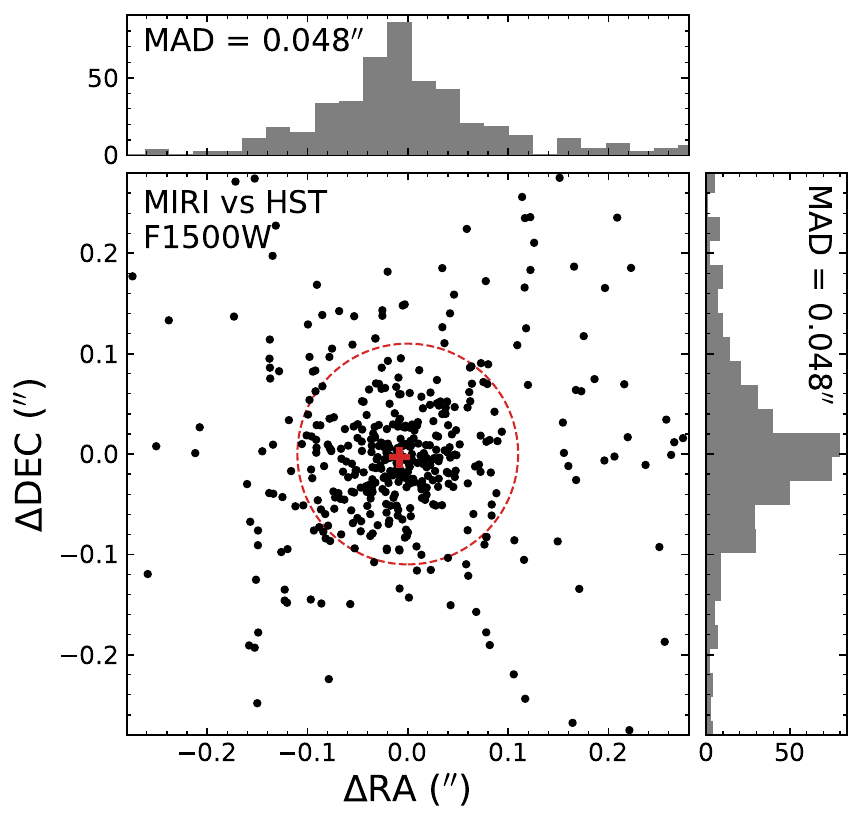}
    \caption{Positional differences between MIRI F560W (left) and F1500W (right) sources and their \textit{HST}\ counterparts. 
    The dashed circle centers at (0,0) and has a radius of the MIRI native pixel scale ($0.11''$).
    The red cross represents the median position of all points. 
    The MIRI sources are detected in the final mosaic. 
    The \textit{HST} astrometry is aligned to \textit{Gaia} DR3. 
    We reach an astrometry accuracy of $\sim 0.023\arcsec$ (about 0.2~pixels) in F560W and $\sim 0.046\arcsec$ (about 0.4~pixels) in F1500W, as labeled. }
    \label{fig:astrometry}
\end{figure*}

The first step is ``TweakReg'' for astrometry correction. 
The correction aims to align MIRI astrometry to the \textit{Gaia}-DR3 astrometric reference frame \fst{\citep{gaiadr3_doi}}. 
However, due to MIRI's relatively small field of view (FOV $\sim 2$~arcmin$^2$), there are typically very few (or even zero) Gaia objects in a MIRI pointing. 
As a result, the default TweakReg fails for all our MIRI data.
We thereby modify the TweakReg raw code to overcome this using a method applied to the CEERS NIRCam data. 
The details are presented in \cite{bagley22}, and we briefly describe the modification here.
Our modified TweakReg aligns each MIRI dither with a \textit{HST}/F160W reference catalog\footnote{\url{http://doi.org/10.17909/T94S3X}} \citep{koekemoer11} in the EGS field, which is itself aligned with \textit{Gaia} DR3 astrometry \citep{gaia22}.
This reference catalog is also used for the astrometry correction in the CEERS NIRCam data reduction \citep{bagley22}. 
This procedure needs a MIRI source list as input, and we again use \sep\ to perform source detection.
The default TweakReg minimizes the root-mean-square error (RMSE) between MIRI and the reference, and our modified version minimizes the median 
\fst{absolute} error (MAE) instead, because MAE is less sensitive to outliers (due to source mismatch or intrinsic positional difference in MIR and NIR).
In our TweakReg run, we set the parameters to \texttt{tolerance = 0.5}, \texttt{searchrad = 1}, and \texttt{separation = 0.01}, which yields an optimal result.
We ran TweakReg twice, as the second iteration slightly improves the result. 
Fig.~\ref{fig:astrometry} displays the RA/DEC differences between MIRI F560W and F1500W sources (detected in the final mosaic) and their \textit{HST}\ counterparts as examples.
The typical RA/DEC dispersion (as measured by median absolute difference, MAD) are $\sim 0.023''\sim 0.2$~pixels (F560W) and $\sim 0.046''\sim 0.4$~pixels (F1500W).

The second and third steps are ``SkyMatch'' and ``Outlier Detection'', respectively.
SkyMatch calculates the sky values (a single value for each dither), and Outlier Detection rejects the remaining cosmic rays that are missed by the Jump step in stage 1 (\S\ref{sec:stage1}).
Even though we have already subtracted the sky background in \S\ref{sec:stage2}, we ran SkyMatch here to ensure the metadata (e.g., FITS-file headers) are set correctly. 
In this step, we set the parameters to \texttt{skymethod = local} and \texttt{subtract = True} (because the median sky is already subtracted to zero in \S\ref{sec:stage2}).
For Outlier Detection, we do not change the default parameters. 

\begin{figure*}[htp]
    \centering
	\includegraphics[width=2\columnwidth]{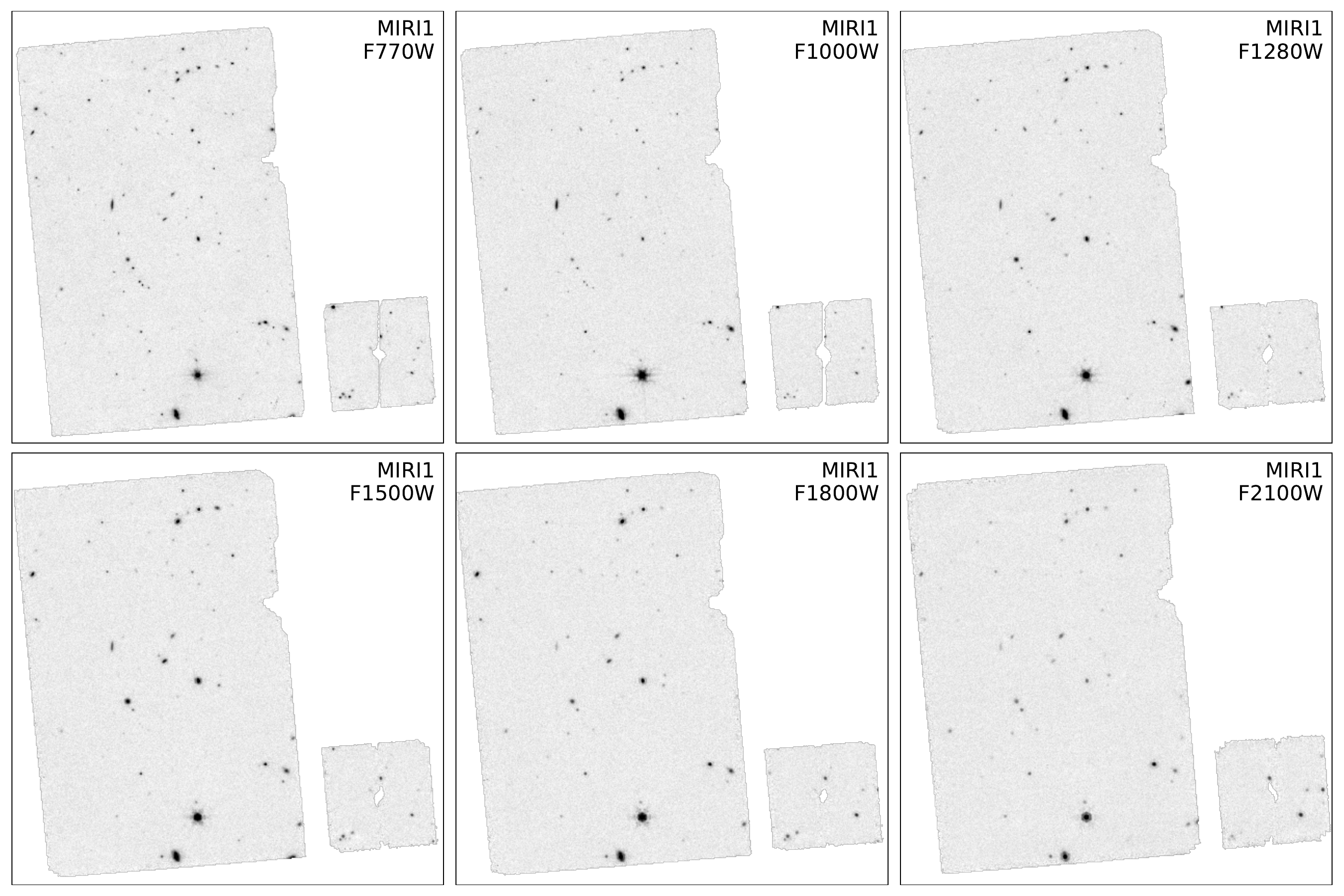}
    \caption{Final mosaics of all six bands in the MIRI1 pointing. 
    The mosaics are astrometrically aligned to Gaia-DR3 as described further in the text of \S\ref{sec:stage3}. 
    See Fig.~\ref{fig:rgb} for an RGB version of a portion of this pointing. 
    }
    \label{fig:miri1_map}
\end{figure*}

The last step is ``Resampling'', which resamples each dither and combines multiple dithers into a final mosaic. 
Following the convention of CEERS \citep{bagley22}, we drizzle the final mosaic into a WCS frame with the same tangent point as the CEERS \textit{HST}\ mosaic.\footnote{ceers.github.io/hdr1.html}
The tangent (RA,DEC) is set by the ``crval'' parameter. 
We also set ``pixel\_scale`` to $0.09''$ (three times that of the \textit{HST} mosaic) and ``rotation $=-49.7$~deg'' (same as the \textit{HST} mosaic). 
These settings ensure that the final MIRI mosaic aligns with the \textit{HST} mosaic, which is the same case for CEERS/NIRCam imaging data products \citep{bagley22}.
This alignment is convenient for some photometry purposes (e.g., \tphot; \citealt{merlin15}).
We set the ``output\_shape'' to $1500\times 1500$ pixels for the Epoch~1 (June) observations (MIRI1, MIRI2, MIRI3, and MIRI6)  and $1800\times 2000$ pixels for the Epoch~2 (December) observations (MIRI5, MIRI7, MIRI8, and MIRI9). 
These frames are sufficiently large to hold all detector areas including the main imager and the additional Lyot-coronagraph region.
The frame for the Epoch~1 pointings is smaller than that for the Epoch~2 pointings, because Epoch~2 observations have larger dithers as needed by the NIRCam/WFSS parallel. 
Fig.~\ref{fig:miri1_map} displays the final mosaics of all MIRI1 bands.

Aside from the science mosaic, the output FITS file also includes an RMS map and a weight map. 
The RMS map accounts for Poisson and readout noises for each pixel, but it does not consider pixel-correlated noise.
Also, compared to the normal detector area, the RMS values are unreasonably much smaller for the areas affected by bad pixels, which we suspect is incorrectly accounted for in the  current \pipeline.
Therefore, we opt to construct a new RMS map for each mosaic.  
To perform this task, we employ \astrorms\footnote{https://github.com/mmechtley/astroRMS} and use the weight map as input.
\astrorms, written by M.\ Mechtley, is a \python\ code that implements the ``acall'' algorithm from \textsc{iraf} (author: M.~Dickinson, private communication).
It calculates an RMS map for an image as ${\rm RMS} = F / \sqrt{\rm weight}$, where the $F$ is a normalization factor (a single value for the mosaic) accounting for pixel-to--pixel correlated noise. 
\astrorms\ estimates the $F$ factor by calculating the autocorrelation function of the mosaic, after masking out sources.     
We use the default parameters of \astrorms.

\section{Quality Assessment}
\label{sec:assess}

\subsection{Imaging Depth}
\label{sec:depth}
We employ the ``ImageDepth'' function of \textsc{photutils} \citep{bradley20} to estimate the 5$\sigma$ depth for each pointing/band. 
This function randomly places 100 circular non-overlapping apertures on each mosaic, avoiding source regions (as indicated by the segmentation map produced by the ``detect\_sources'' function of \textsc{photutils}). 
Here, we adopt an aperture radius equal to the full width at half maximum (FWHM).
The FWHM values are $0.207\arcsec$ (F560W), $0.269\arcsec$ (F770W), $0.328\arcsec$ (F1000W), $0.420\arcsec$ (F1280W), $0.488\arcsec$ (F1500W), $0.591\arcsec$ (F1800W), and $0.674\arcsec$ (F2100W) according to the \textit{JWST} User Documentation.\footnote{https://jwst-docs.stsci.edu/jwst-mid-infrared-instrument/miri-performance/miri-point-spread-functions}
ImageDepth then calculates the standard deviation of the fluxes encircled by these random apertures, and multiplies this value by the significance level (i.e., 5 in our case) as the 5$\sigma$ limiting flux. 
To correct the aperture flux to total, we applied a filter-dependent correction to each limiting flux.
This aperture-correction factor is estimated from the encircled energy fractions (EEF) versus radius curve, which is derived based on \textsc{webbpsf} \fst{\citep{webbpsf15}} with the latest Project Reference Database (PRD) version, PRDOPSSOC-059. 

The results are listed in Table~\ref{tab:depth}. 
Table~\ref{tab:depth} also compares the predicted 5$\sigma$ depth from the ETC.  
In the ETC calculation, we adopt the same aperture radius (i.e., $R=\rm FWHM$) as in our depth estimation for the real imaging data.
For all bands except F2100W, our measured depths are similar to the corresponding ETC predicted ones with a difference within $\sim \pm 0.4$~mag, despite that ETC does not consider the effects of dithering.
For F2100W, our measured depth is shallower than the ETC prediction by $\sim 0.9$~mag.
This relatively large difference could be caused by the excess noise in this band compared to the ETC expectations (see Fig.~\ref{fig:bkg}). 



\begin{table*}
\centering
\caption{Measured 5$\sigma$ depth in AB mag (and ETC predictions)}
\small
\label{tab:depth}
\begin{tabular}{cccccccc} \hline\hline
Field & F560W & F770W & F1000W & F1280W & F1500W & F1800W & F2100W \\
\hline 
MIRI1  & n/a & 25.59 (25.23) & 24.84 (24.67) & 24.17 (24.05) & 23.64 (23.76) & 22.90 (22.96) & 22.20 (23.07) \\ 
MIRI2  & n/a & 25.58 (25.23) & 24.81 (24.67) & 24.12 (24.05) & 23.75 (23.76) & 22.81 (22.96) & 22.43 (23.35) \\ 
MIRI3  & 26.19 (26.17) & 26.47 (26.09) & n/a & n/a & n/a & n/a & n/a \\ 
MIRI6  & 26.27 (26.17) & 26.42 (26.09) & n/a & n/a & n/a & n/a & n/a \\ 
MIRI5  & n/a & n/a & 24.66 (24.52) & 23.64 (23.73) & 22.99 (23.43) & 22.40 (22.74) & n/a \\ 
MIRI8  & n/a & n/a & 24.71 (24.52) & 23.60 (23.73) & 23.15 (23.43) & 22.40 (22.74) & n/a \\ 
MIRI7  & 25.85 (25.74) & 25.32 (25.44) & n/a & n/a & n/a & n/a & n/a \\ 
MIRI9  & 25.93 (25.74) & 25.40 (25.44) & n/a & n/a & n/a & n/a & n/a \\ 
\hline
\end{tabular}
\begin{flushleft}
{\sc Note.} --- The depth is measured with an aperture radius of $R=\rm FWHM$ for each pointing/band, and an aperture correction has been applied (assuming a point source). ``n/a'' indicates the filter is not available in the pointing. 
\end{flushleft}
\end{table*}


\subsection{Comparison of MIRI and \textit{Spitzer} photometry at 5.6 and 7.7 $\mu$m}
\label{sec:spitzer}
The MIRI F560W and F770W filters are similar to the \textit{Spitzer}/Infrared Array Camera (IRAC) Ch3 (5.8~$\mu$m) and Ch4 (8.0$\mu$m) filters, and we can thereby take advantage of the well-established IRAC photometry \citep{stefanon17} in the CEERS field to evaluate the imaging data quality of our MIRI F560W and F770W.
To perform this task, we use the four blue pointings of MIRI3, MIRI6, MIRI7, and MIRI9, each having both F560W and F770W.   

Our photometry extraction here is similar to the method in \cite{papovich23}.
In this method, a detection map from F560W and F770W for each pointing is created first. 
To do this, we generate the \fst{point spread function (PSF)} for F560W and F770W, respectively, utilizing \textsc{webbpsf}.
We then employ \textsc{pypher} \citep{boucaud16} to construct a PSF convolution kernel from F560W to F770W.   
After convolving with this kernel, the F560W PSF becomes similar to the F770W PSF: the difference between their EEF is $\lesssim 3\%$ for any radius. 
We convolve our F560W mosaics with this kernel, resulting in images with ``PSF matched'' to F770W. 
Finally, we add up the PSF-matched F560W and F770W images , weighted by their corresponding weight maps (\S\ref{sec:stage3}), and adopt the result as the detection image.   
In addition, we also build a weight map for the detection image by adding up the weight maps of F560W and F770W.

We ran \se\ (version 2.19.5; \citealt{bertin96}) in the ``dual-image'' mode, where we use the F560W$+$F770W detection image and its associated weight map for source detection. 
For flux measurements, we use the PSF-matched F560W and F770W images and their associated RMS maps (\S\ref{sec:stage3}). 
The parameters are detailed in Table~1 of \cite{papovich23}.
Briefly, the key parameters are \texttt{DETECT\_MINAREA}$=$ 10 pixels, \texttt{DETECT\_THRESH}$=1.3$, and \texttt{FILTER\_NAME}$=$ gauss\_2.5\_5$\times$5. 
From the \se\ output, we adopt the \texttt{FLUX\_AUTO} as our measured total fluxes. 
To avoid false detections, we filter out sources with $\rm S/N<1$ in both F560W and F770W fluxes.
The source number per pointing ranges from $\sim 260$--310.
\fst{We publicly release our MIRI photometry catalog at the CEERS website (\url{https://ceers.github.io/}).
Table~\ref{tab:cat} shows the format of this catalog. 
}

\begin{figure*}[htp]
    \centering
	\includegraphics[width=1.6\columnwidth]{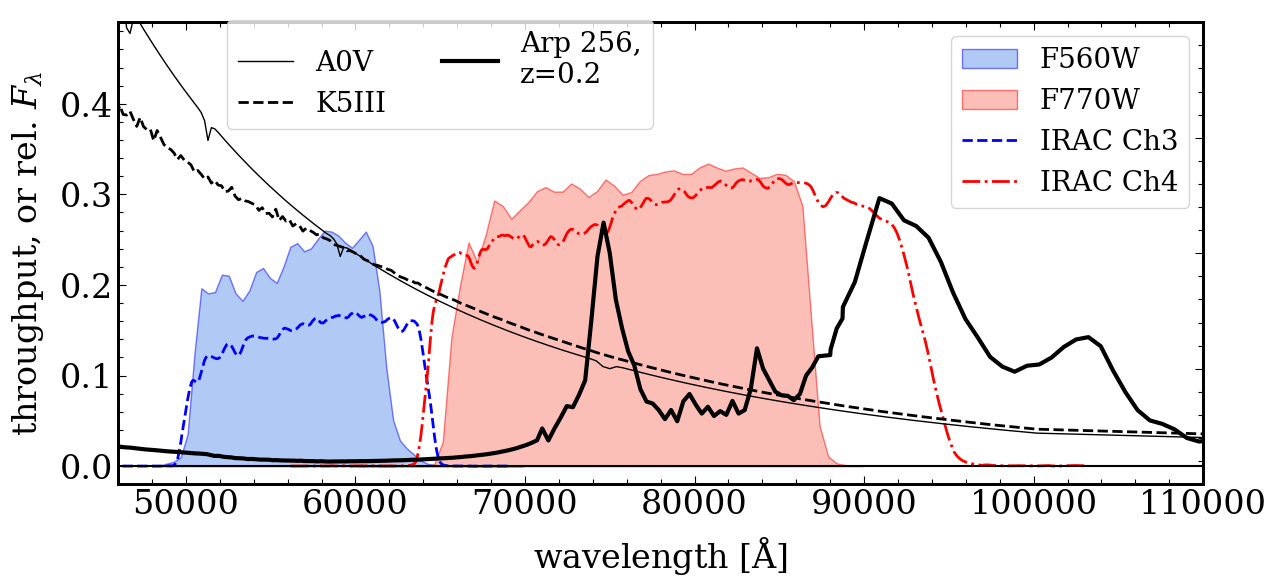}
    \caption{Comparison of MIRI versus IRAC filters. 
    The blue/red shaded regions and the blue/red curves indicate the MIRI F560W/F770W and IRAC Ch3/Ch4, respectively. 
    The black thick solid curve represents the SED of Arp~256 \citep{brown14}, redshifted assuming $z=0.2$. 
    The black thin solid and dashed curves represent the SEDs of A0V and K5III stars, respectively. 
    }
    \label{fig:IRAC_MIRI_filters}
\end{figure*}

\begin{figure*}[htp]
    \centering
	\includegraphics[width=2\columnwidth]{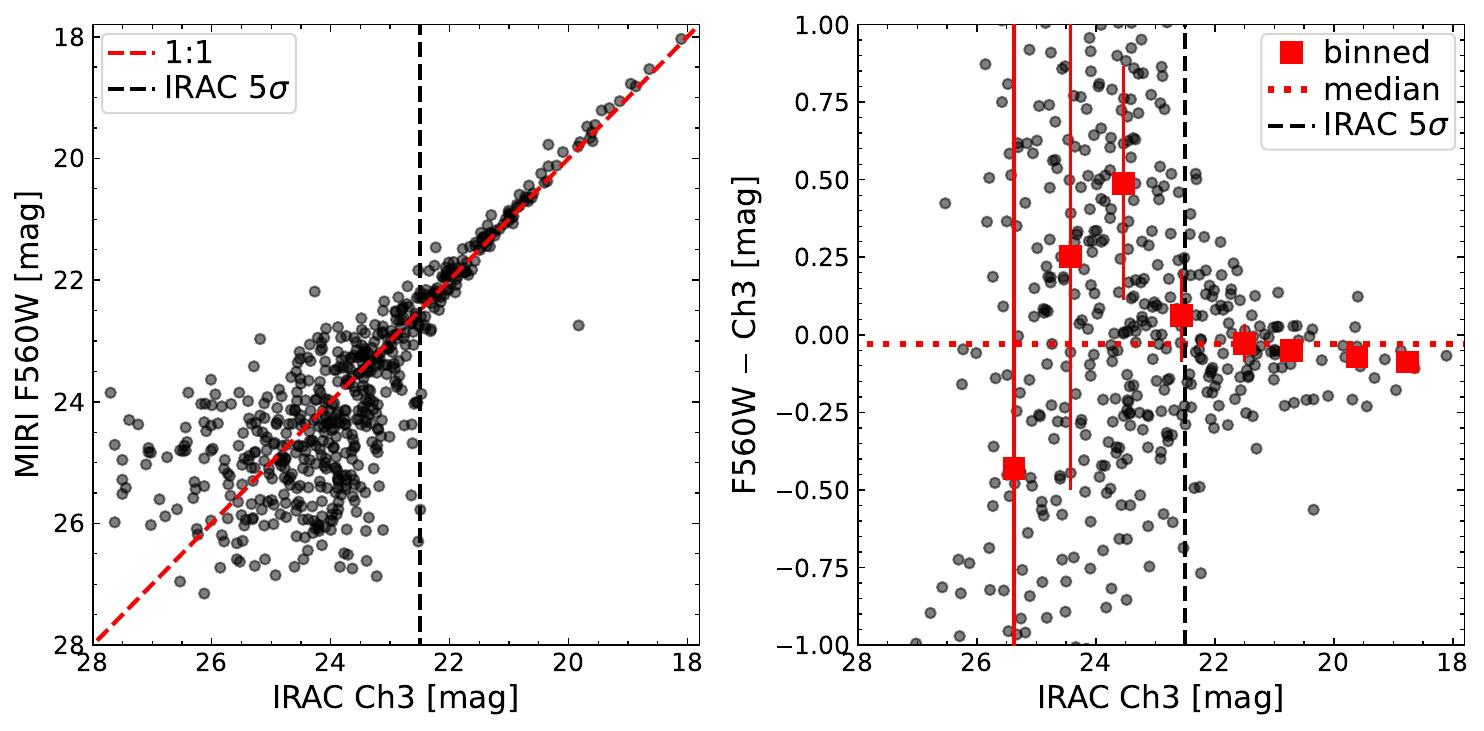}
	\includegraphics[width=2\columnwidth]{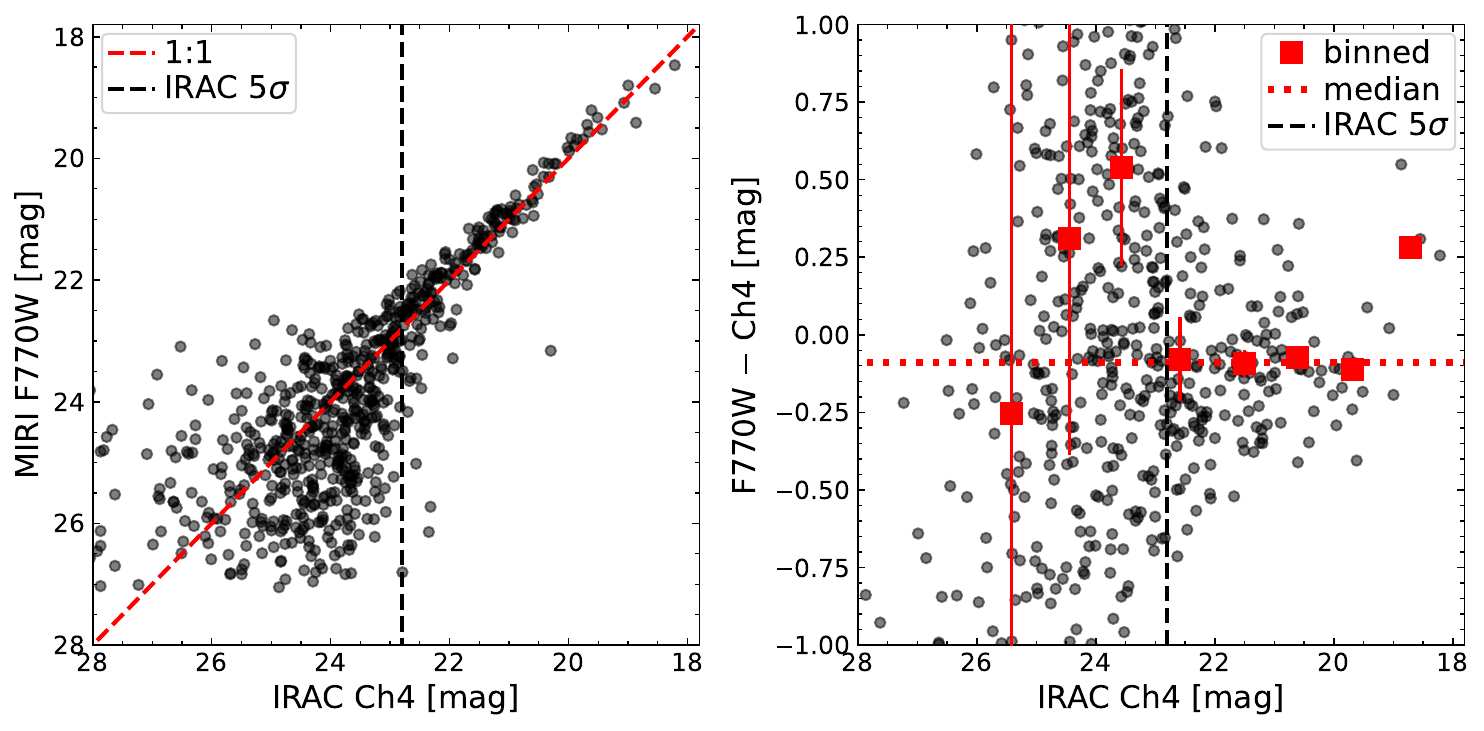}
    \caption{Comparison of source magnitudes measured in MIRI F560W and F770W versus IRAC Ch3 and Ch4.
    The top panels compare MIRI F560W versus IRAC Ch3, and the bottom panels compare MIRI F770W versus IRAC Ch4. 
    The red dashed lines indicate the 1:1 (unity) relation.  
    The black vertical lines show the $5\sigma$ detection limits for the IRAC data.     
    The right panels display the difference between the two magnitudes.
    The red dotted lines indicate the median mag difference of bright sources (IRAC $<23$~mag). 
    Each large red squares represent the median offset of sources in each magnitude bin, and its error bar represents the median errors of the sources in the bin.
    %
    }
    \label{fig:miri_vs_irac}
\end{figure*}

Fig.~\ref{fig:miri_vs_irac} compares the resulting MIRI fluxes versus the IRAC fluxes. 
We calculate the median offset based on sources brighter than 23~mag in the IRAC band.
The median offsets are $-0.03$~mag (F560W $-$ Ch3) and $-0.09$~mag (F770W $-$ Ch4), as indicated by the red dotted horizontal lines.
We consider such level of offsets ($<0.1$~mag) are acceptable, as the MIRI and IRAC filter transmissions are not identical (see Fig.~\ref{fig:IRAC_MIRI_filters}) and the MIRI and IRAC photometry are calibrated independently.  
To check the magnitude dependence of the offset,
we also bin the sources by their IRAC magnitudes and calculate the median $\Delta$mag for each bin (large red squares in Fig.~\ref{fig:miri_vs_irac}).
These median $\Delta$mag are generally flat as a function of magnitude, except for the brightest bin in F770W $-$ Ch4 (for which F770W is fainter than Ch4).
Upon inspecting the outliers in this bin, we attribute the causes to source confusion in Ch4, and fact that the IRAC Ch4 filter extends to longer wavelengths than MIRI F770W (this means that the redshifted PAH emission from galaxies can lead to ``$K$--corrections'' between magnitudes measured in the different bands, see e.g, the galaxy SED in Fig.~\ref{fig:IRAC_MIRI_filters}).

\begin{figure}[htp]
    \centering
    \includegraphics[width=\columnwidth]{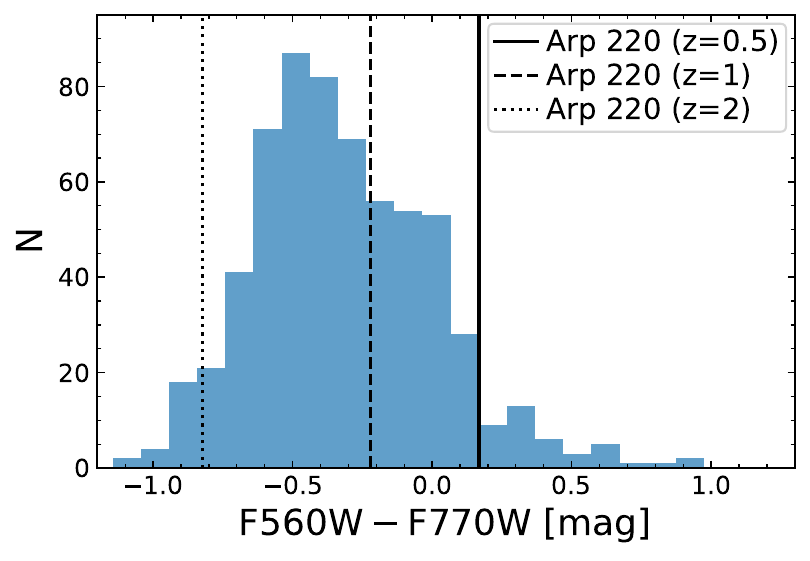}
    \caption{F560W $-$ F770W color distribution. 
    Here, for accuracy, we only adopt sources with $\rm S/N>3$ in both bands.
    The vertical black solid, dashed, and dotted lines indicate the colors of Arp~220 at $z=0.5$, $z=1$, and $z=2$, respectively \citep{brown14}.
    The bulk ($84\%$) of sources have negative color values, as expected from starlight in higher redshift galaxies $(z \gtrsim 1)$ .  
    }
    \label{fig:color_hist}
\end{figure}

Fig.~\ref{fig:color_hist} displays the F560W $-$ F770W color distribution.
The bulk ($84\%$) of sources have negative color values, as expected from starlight that typically declines toward longer wavelengths at $\gtrsim 1\ \mu$m (rest-frame).

\begin{table*}
\centering
\caption{\fst{MIRI photometry catalog}}
\small
\label{tab:cat}
\begin{tabular}{ccccccccccc} \hline\hline
ID & RA & Dec & Pointing & F560W & F560W\_err & F770W & F770W\_err & ... & F2100W & F2100W\_err \\
\hline 
ceers\_miri\_0001 & 214.708041 & 52.765937 & miri9 & nan & nan & 3.728 & 0.322 &  ...  & nan & nan \\ 
ceers\_miri\_0002 & 214.708355 & 52.766147 & miri9 & nan & nan & 2.538 & 0.295 &  ...  & nan & nan \\ 
ceers\_miri\_0003 & 214.709729 & 52.762546 & miri9 & nan & nan & 0.936 & 0.187 &  ...  & nan & nan \\ 
ceers\_miri\_0004 & 214.713093 & 52.758558 & miri9 & nan & nan & 2.776 & 0.304 &  ...  & nan & nan \\ 
ceers\_miri\_0005 & 214.713695 & 52.763276 & miri9 & nan & nan & 5.392 & 0.516 &  ...  & nan & nan \\ 
ceers\_miri\_0006 & 214.714152 & 52.757493 & miri9 & nan & nan & 3.924 & 0.37 &  ...  & nan & nan \\ 
ceers\_miri\_0007 & 214.714153 & 52.769189 & miri9 & nan & nan & 42.768 & 0.516 &  ...  & nan & nan \\ 
ceers\_miri\_0008 & 214.715846 & 52.769588 & miri9 & nan & nan & 3.249 & 0.327 &  ...  & nan & nan \\ 
ceers\_miri\_0009 & 214.716106 & 52.756932 & miri9 & 1.548 & 0.259 & 1.365 & 0.166 &  ...  & nan & nan \\ 
ceers\_miri\_0010 & 214.716621 & 52.764468 & miri9 & nan & nan & 0.465 & 0.18 &  ...  & nan & nan \\ 
\hline
\end{tabular}
\begin{flushleft}
\fst{ {\sc Note.} --- Only a portion of this table is shown here, and the full version is available at the CEERS website. 
The table is sorted in the ascending order of RA.
Fluxes and their errors are in the units of $\mu$Jy (``nan'' means the band is not available for the source). 
They are extracted using either \se\ (blue pointings; 
 \S\ref{sec:spitzer}) or \tphot\ (red pointings; \S\ref{sec:star}). }
\end{flushleft}
\end{table*}

\subsection{MIRI SEDs of Galactic stars}
\label{sec:star}
In \S\ref{sec:spitzer}, we have evaluated the data quality of the two blue bands, F560W and F770W, by comparing their fluxes with the well-established IRAC fluxes. 
However, for the redder MIRI bands, there is no previous data with similar band coverage as references. 
Instead, we perform the quality assessment by inspecting the MIRI photometry of the Galactic stars in the red MIRI pointings (MIRI1, MIRI2, MIRI5, and MIRI8), as their MIR SEDs should typically follow the Rayleigh-Jeans law ($F_\nu \propto\ \lambda^{-2}$; e.g., Fig.~\ref{fig:IRAC_MIRI_filters}).


For photometry of sources in the red MIRI pointings, we ran \tphot\ \citep[version 2.1][]{merlin15} for photometry extraction following \cite{yang21}.   
We used \tphot\ rather than the method adopted for the blue pointings (see \S\ref{sec:spitzer}), because, e.g., smoothing F770W to the F2100W PSF level would significantly degrade the F770W imaging quality.  
\tphot\ convolves the source cutouts from a ``high-resolution'' image with a convolution kernel, and normalize the total flux of each cutout to unity. 
These normalized cutouts are called ``templates''. 
It then fits these templates to the ``low-resolution'' image, adjusting the normalization of each template (and an additional background) to find the least-$\chi^2$ solution.
The normalization of each template is considered as the best-fit fluxes for the corresponding source. 
In our run, we used the \textit{HST}/F160W with a pixel scale of $0.03''/\rm pixel$ (see \S\ref{sec:stage3}) as the high-resolution image and MIRI as the low-resolution images.
\textit{HST}/F160W is adopted because it fully covers all of our MIRI pointings.  
The convolution kernels from F160W to MIRI bands are constructed using the same method as in \S\ref{sec:spitzer}.
In the future, we also plan to extract the MIRI photometry using CEERS/NIRCam as the high-resolution images for the MIRI-NIRCam overlapping areas (but as noted above, NIRCam imaging is not available over all of the CEERS red MIRI pointings). 

\tphot\ also needs a catalog and segmentation map derived from the high-resolution image to build the source templates. 
To generate these inputs, we ran \se\ with parameters of \texttt{DETECT\_MINAREA}$=$ 10 pixels, \texttt{DETECT\_THRESH}$=0.7$, and \texttt{FILTER\_NAME}$=$ gauss\_9.0\_9$\times$9.
In the \tphot\ output, we remove a flux entry if it has $\rm S/N<1$, which indicates the template normalization cannot be tightly constrained in the fit. 
We remove a source if it does not have any valid flux over all the MIRI bands.
In the final clean catalog, the source number per pointing ranges from $\sim 360$--600.
\fst{The \tphot\ photometry is also included in our data release (Table~\ref{tab:cat}).}

\begin{figure*}[htp]
    \centering
	\includegraphics[width=0.75\columnwidth]{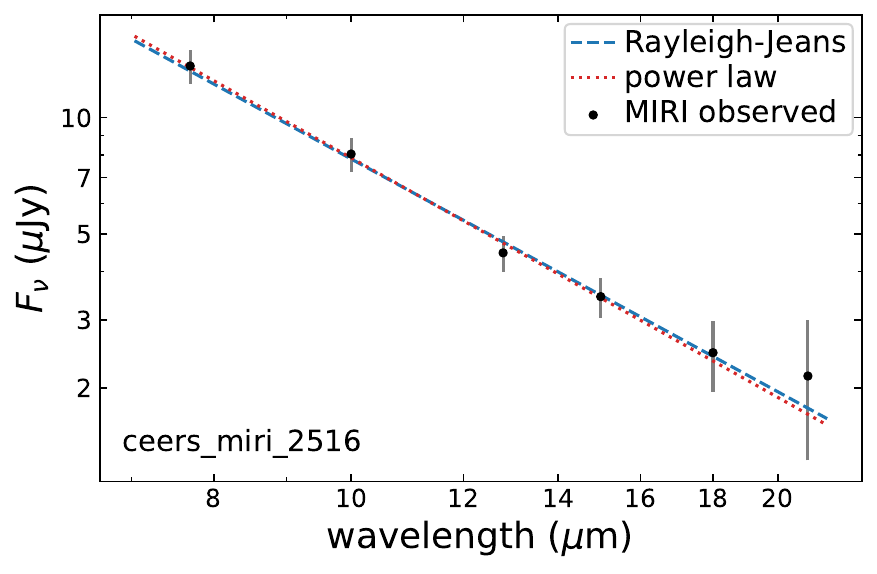}
	\includegraphics[width=1.0\columnwidth]{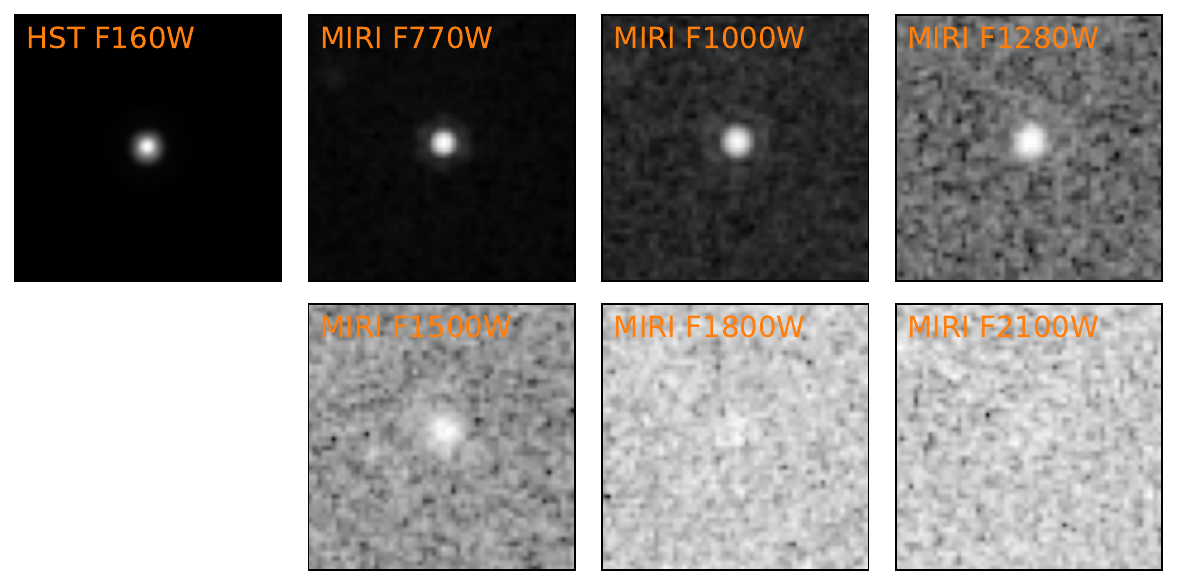}
 	\includegraphics[width=0.75\columnwidth]{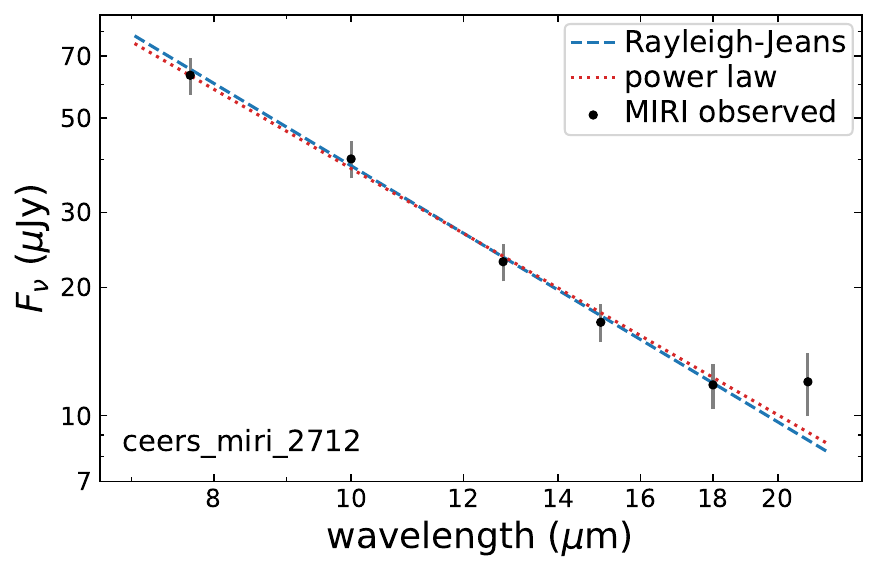}
	\includegraphics[width=1.0\columnwidth]{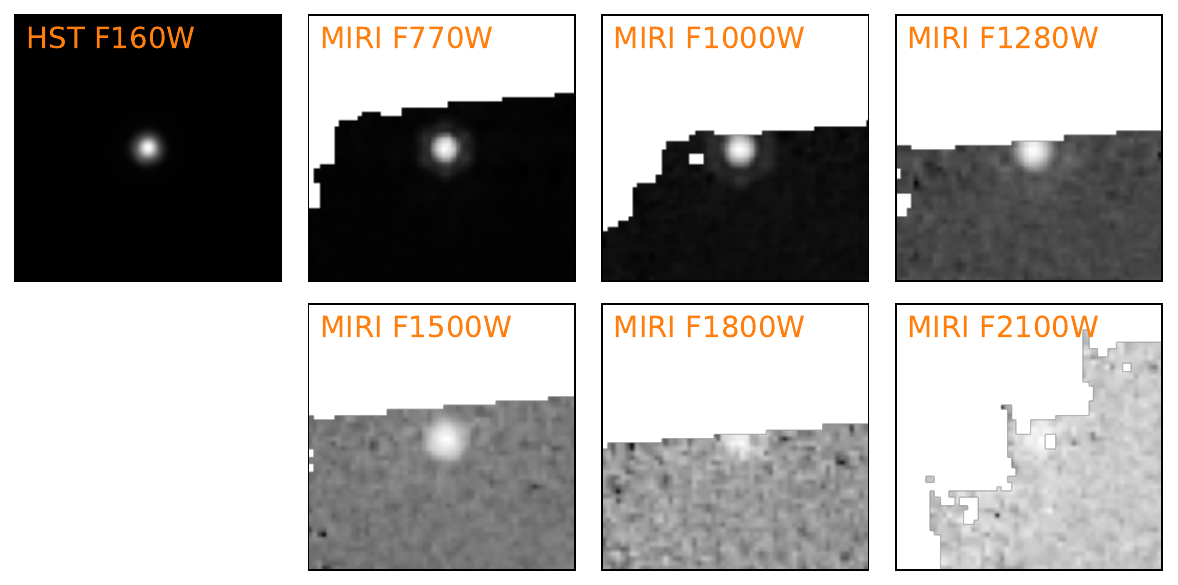}
 	\includegraphics[width=0.75\columnwidth]{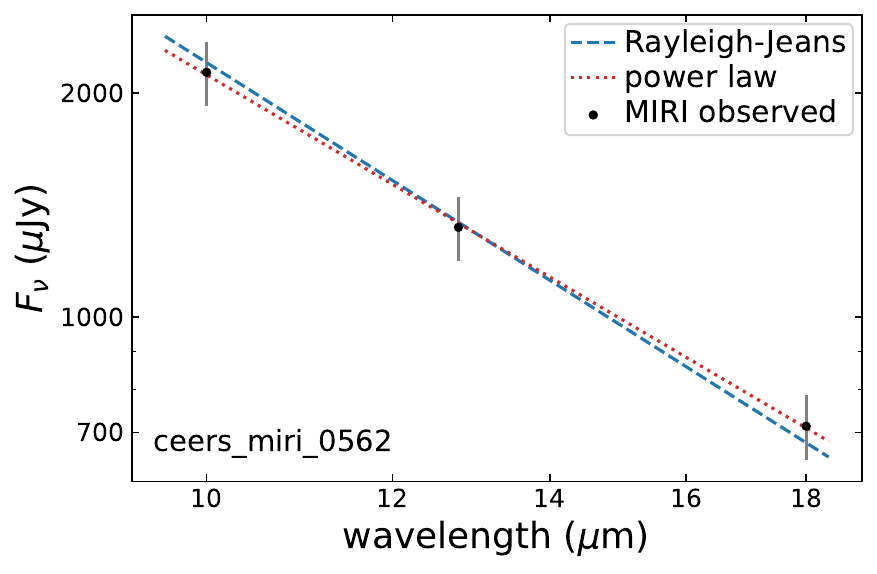}
	\includegraphics[width=1.0\columnwidth]{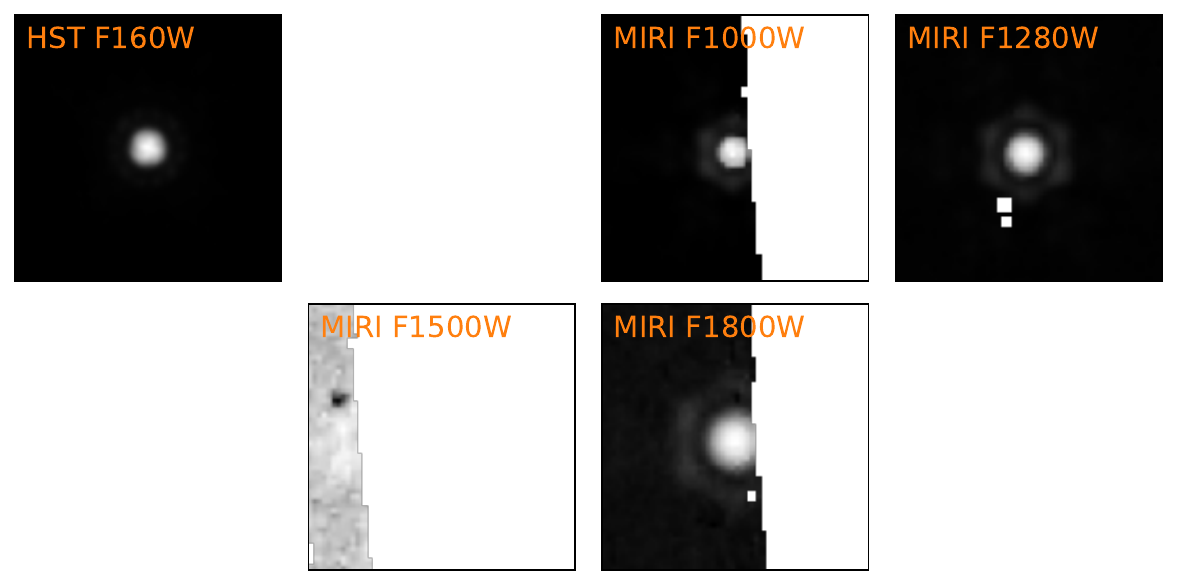}
 	\includegraphics[width=0.75\columnwidth]{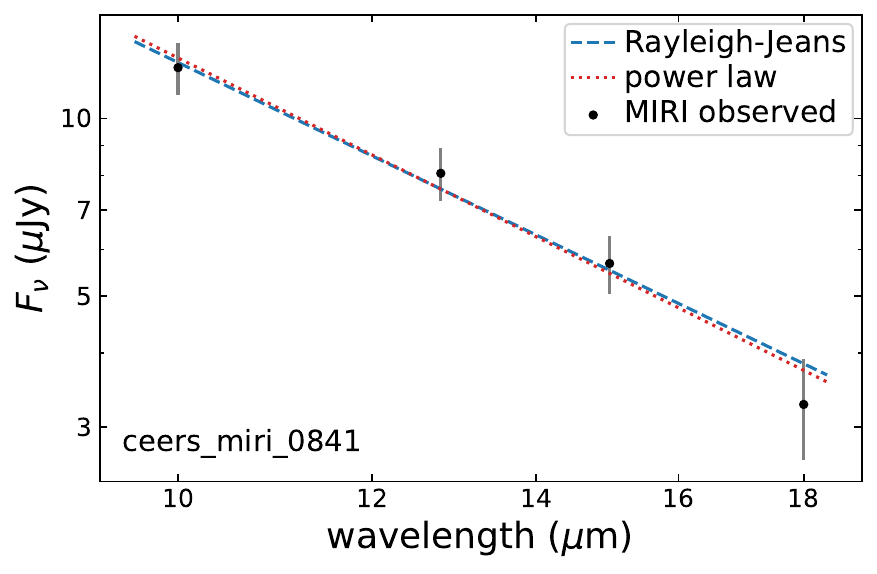}
	\includegraphics[width=1.0\columnwidth]{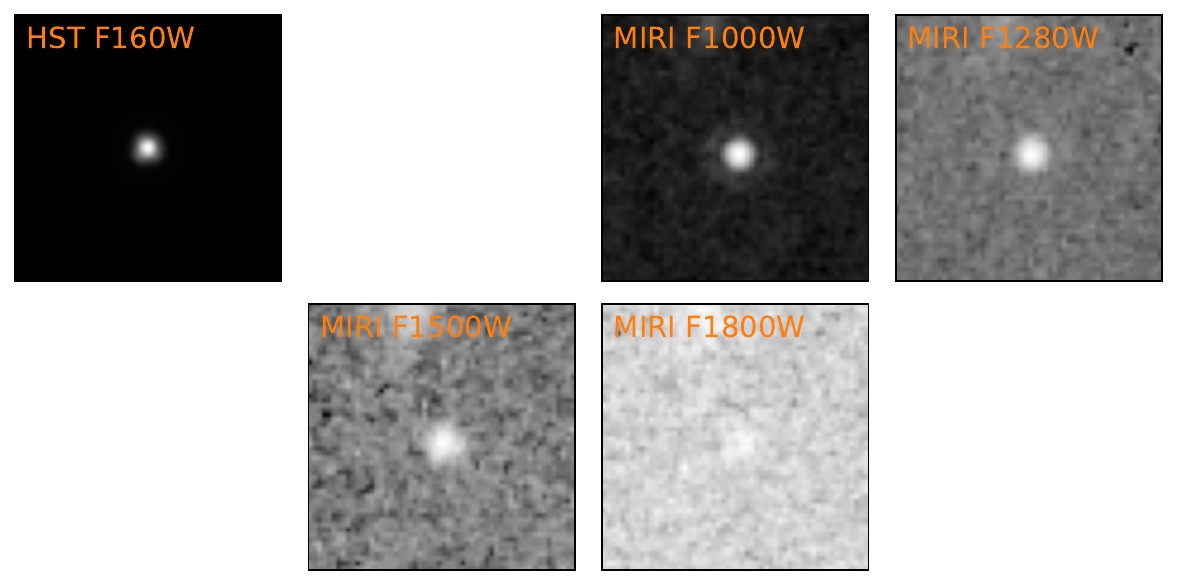}
 	\includegraphics[width=0.75\columnwidth]{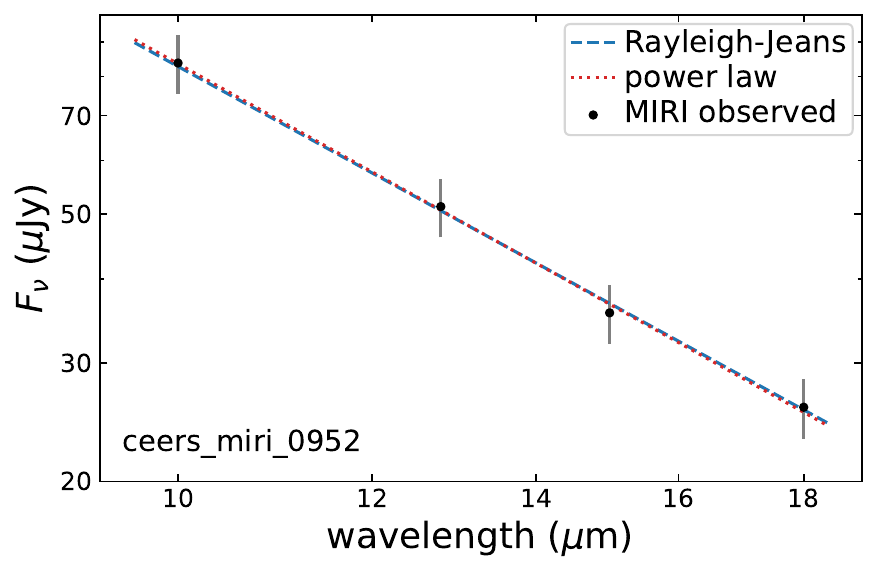}
	\includegraphics[width=1.0\columnwidth]{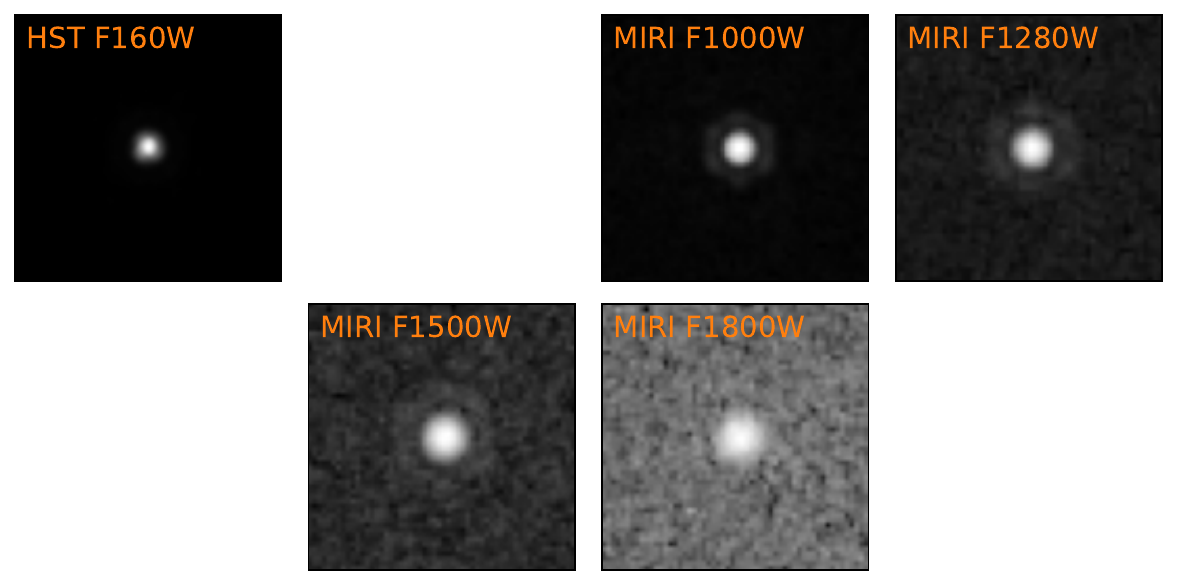}
    \caption{\textit{Left}: MIRI SED (left) of the Galactic stars covered by MIRI1.
    The blue dashed and red dotted lines represent the best fits of the Rayleigh-Jeans law (normalization as free parameter) and a power law (normalization and index as free parameters).
    The two fits are very similar, indicating our MIRI colors are reliable.
    \textit{Right}: The corresponding $5\times 5 ''$ cutouts of \textit{HST}/F160W and MIRI.
    Although the sources are sometimes truncated by the detector edge (e.g., F1800W of the 2nd  star, from top to bottom), the corresponding photometry is still acceptable.
    This shows that, for these sources, \tphot\ can still robustly constrain the flux by fitting the pixels within the detector coverage.  
    }
    \label{fig:star_sed}
\end{figure*}

We match our MIRI sources versus the robust Galactic stars ($\rm S/N>3$ parallax and/or proper motion measurements) in \textit{Gaia} DR3 \fst{\citep{gaiadr3_doi}}, using a $0.5''$ matching radius, resulting in five matches.
The MIRI SEDs of these stars are displayed in Fig.~\ref{fig:star_sed}. 
Following the convention of \textsc{cigale} \citep{boquien19}, we add a 10\% uncertainty in quadrature with the \tphot\ error to account for, e.g., calibration uncertainties.
This uncertainty floor also prevents the bluer bands (which often have higher S/N) from being over-weighted in the fit. 
We first fit the MIRI data points with the Rayleigh-Jeans law, using \textsc{scipy.optimize.curve\_fit} (blue dashed line in Fig.~\ref{fig:star_sed}).
We then fit the points with a power-law model (normalization and index as free parameters; red dotted line in Fig.~\ref{fig:star_sed}).

The best-fit models are displayed in Fig.~\ref{fig:star_sed}.
The best-fit power-law indexes are consistent with the Rayleigh-Jeans model (index $=2$) at a 68\% confidence level.
The similarity of the two fits demonstrates that our MIRI colors are reliable in the red pointings of MIRI1, MIRI2, MIRI5, and MIRI8.
Quantitatively, the MAD of our MIRI photometry versus the Rayleigh-Jeans model is only $0.03$~mag.
We highlight that, although the sources are sometimes truncated by the detector edge (e.g., F1800W of the 2nd star, from top to bottom), the resulting photometry is still acceptable.
This shows that, for these sources, \tphot\ can still robustly constrain the flux by fitting the pixels within the detector coverage.  


\subsection{MIRI number counts}
\label{sec:cnts}
\begin{figure*}[htp]
    \centering
    \includegraphics[width=1.8\columnwidth]{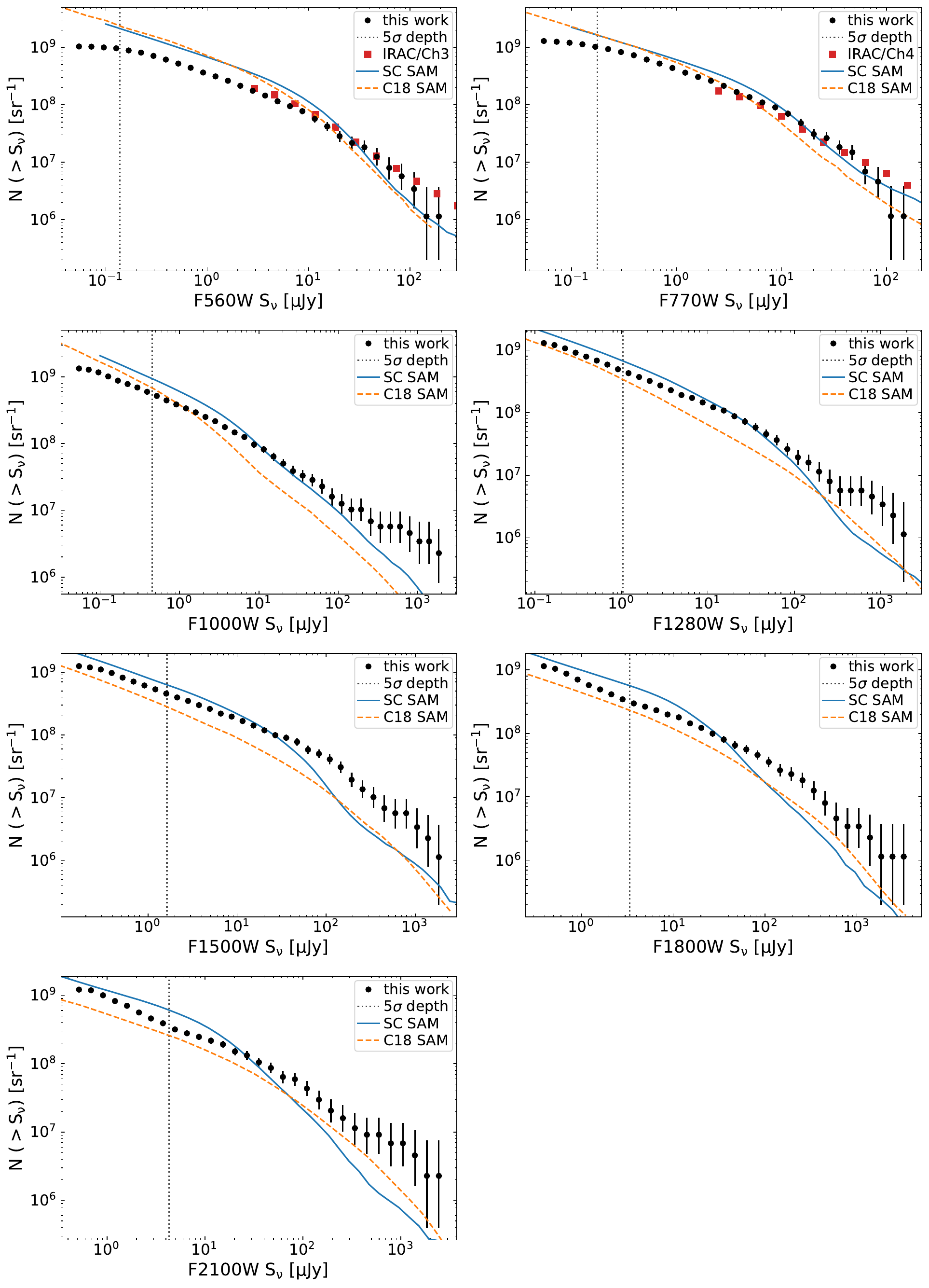}
    \caption{Cumulative number counts of different MIRI bands. 
    The error bars represent Poisson uncertainties calculated using \textsc{astropy.stats.poisson\_conf\_interval}. 
    The dashed vertical line in each panel indicates the median 5$\sigma$ depth of the CEERS pointings (Table~\ref{tab:depth}).
    The red squares in the F560W/F770W panel indicate the IRAC Ch3/Ch4 number counts from \cite{fazio04}.
    The blue solid and orange dashed curves represent the predictions of the Santa Cruz and \cite{cowley18} semi-analytic models (SAMs), respectively. 
    The F560W and F770W number are based on \se\ photometry from the blue MIRI pointings (MIRI3, MIRI6, MIRI7, and MIRI9), while the F1000W, F1280W, F1500W, F1800W, and F2100W are based on \tphot\ photometry from the red MIRI pointings  (MIRI1, MIRI2, MIRI5, and MIRI8).
    All of the MIRI number counts and their associated errors are publicly available with this figure.
    }
    \label{fig:logN_logS}
\end{figure*}

Using the extracted photometry (\S\ref{sec:spitzer} and \S\ref{sec:star}), we derive the cumulative number counts and display the results in Fig.~\ref{fig:logN_logS}.
The uncertainties are estimated with Poisson errors calculated using \textsc{astropy.stats.poisson\_conf\_interval}.
These number counts and their associated errors are publicly available in an electronic table associated with this paper.

We over-plot IRAC number counts \citep[in the panels of F560W and F770W][]{fazio04}.
The IRAC number counts are similar to the MIRI values as expected, and we attribute the minor difference to the different filter shapes of MIRI versus IRAC (see Fig.~\ref{fig:IRAC_MIRI_filters}) and their calibration uncertainties.

\fst{The number counts should be dominantly contributed by extragalactic sources, because there are only five Galactic stars (identified by \textit{Gaia}; \S\ref{sec:star}) in our MIRI data.
Therefore, we can compare the observed number counts with predictions from extragalactic models. 
}
Fig.~\ref{fig:logN_logS} also shows the predictions of the cumulative number counts from two different semi-analytic models: the GALFORM model presented in \citet{cowley18} and the Santa Cruz semi-analytic models \citep{somerville12,somerville2015,yung19,somerville2021,yung22}. 
Both models adopt similar recipes for physical processes such as cooling, star formation, stellar feedback, chemical evolution, and black hole growth and feedback, implemented within cosmological merger trees, though these recipes differ in their details. Both models have also been extensively calibrated and shown to produce good agreement with UV-optical galaxy luminosity functions at $z<10$. 
\citet{cowley18} model dust emission by coupling the semi-analytic model GALFORM with the GRASIL dust+radiative transfer models of \citet{silva98}. 
These models adopt an idealized geometry (bulge+disk) and assume a composition and geometry for interstellar dust.  GRASIL then computes the dust emission SEDs of the model galaxies, accounting for absorption and scattering of radiation by dust and its re-emission. 
The Santa Cruz SAMs adopt a simple recipe for dust attenuation as a function of gas column density, metallicity, and redshift \citep{somerville12,somerville2021,yung19}, and then assume that all the energy absorbed by dust is re-radiated in the IR. They adopt the dust emission SED templates of \cite{charyelbaz2001}. 

At the shortest MIRI wavelength (F560W), the model predictions agree well with one another, and are in moderately good agreement with the observations, at the level of a factor of a few. Both models tend to overpredict fainter galaxies and underpredict brighter objects. At F770W, the Santa Cruz models predict more bright objects than the \citet{cowley18} models, and are in better agreement with the observed counts, though both models still slightly overpredict fainter objects. At longer wavelengths (F1000W and above), the Santa Cruz models systematically predict higher counts of faint objects than the \citet{cowley18} models, and agree with or slightly overpredict the observed counts within about a factor of two below a flux of 20-30 $\mu$Jy. The \citet{cowley18} models underpredict the observed counts at all fluxes above about 1 $\mu$Jy. At even longer wavelengths (F1280W and higher), both models produce similar predictions for the brightest counts ($\gtrsim 100$ $\mu$Jy), and both models underproduce the number of objects at these fluxes by up to an order of magnitude. 

Given the complexity of modeling the SED in this wavelength range, and the relative simplicity of these modeling approaches, the agreement of the models with one another (within about a factor of two everywhere) and with the observations (within an order of magnitude) is actually encouraging. Semi-analytic models and hydrodynamic simulations have traditionally found it challenging to reproduce the counts of luminous dusty galaxies at longer wavelengths (e.g., FIR and sub-mm, see the discussion in \citet{cowley18} and references therein, as well as \citet{somerville12}, \citet{niemi12}, \citet{hayward21}). 
As discussed in these works, reasons for this may include a non-universal stellar initial mass function, the modeling of starbursts and dust attenuation, evolving dust temperatures, and quenching by AGN feedback. All of these factors may be important here as well. In addition, galaxy SEDs in the 5--25~$\mu$m range probed by MIRI are strongly affected by PAH absorption and emission features that are notoriously challenging to model. Futhermore, dust heating by accreting SMBHs, which are not directly included as heating sources in either of the models shown here, may be important. It is particularly difficult to disentangle the origin of the discrepancies between the two models, or between the models and the observations, when considering counts that are integrated over all redshifts, and it is beyond the scope of this work to do so. 
However, this comparison highlights the power of the new \textit{JWST} MIRI observations to help constrain and improve physics-based models of galaxy formation.

\section{Summary}
\label{sec:sum}
We have presented our methodology for processing the MIRI imaging data of CEERS, and validated that accurate photometry can be obtained. 
These MIRI imaging have a total of 8 pointings (Table~\ref{tab:obs}).
We have reduced the data using the \textsc{jwst Calibration Pipeline} with our modifications and additional steps designed to enhance the output quality, e.g., improving astrometry and removing detector artefacts (\S\ref{sec:reduction}).
We have estimated the imaging depth for each pointing/band (\S\ref{sec:depth}).
We have extracted MIRI photometry using \se\ and \tphot\ for the blue and red pointings, respectively, and derived number counts of each MIRI band (\S\ref{sec:spitzer} and \S\ref{sec:star}).
We have assessed the imaging quality by comparing the extracted MIRI photometry versus IRAC photometry and investigating the MIRI SEDs of Galactic stars. 
We also derive MIRI number counts and compare them with the IRAC result and theoretical predictions.   
All of these assessments indicate our reduced MIRI photometry has good quality and is useful to our understanding of galaxy evolution. 

We release our data products including reduced mosaics and catalogs at the CEERS official website (\url{https://ceers.github.io/}) and on MAST as High Level Science Products via \dataset[10.17909/z7p0-8481]{\doi{10.17909/z7p0-8481}}.
This release represents one of the first public MIRI surveys. 
Our dataset has already enabled some pioneering studies, covering topics of, e.g., high-redshift AGN, galaxy morphology, and early galaxy evolution (e.g., \citealt{barro23, kirkpatrick23, larson23, magnelli23, papovich23, shen23, yang23}).
In the future, researchers can explore their science topics with our reduced MIRI data products of CEERS. 
They may also benefit from our reduction methods detailed in this work when reducing their own MIRI data.  

\hfill \break

\section*{Acknowledgments}
We thank the referee for helpful feedback that improved this work.
We thank the STScI \textit{JWST} helpdesk (especially Karl Gordon, Mattia Libralato, and Jane Morrison) for suggestions about MIRI data reduction. 
Support for program No.~JWST-ERS01345 was provided through a grant from the STScI under NASA contract NAS5-03127.
GY, KIC and EI acknowledge funding from the Netherlands Research School for Astronomy (NOVA). 
KIC and VK acknowledge funding from the Dutch Research Council (NWO) through the award of the Vici Grant VI.C.212.036.  
Some of the data presented in this paper were obtained from the Mikulski Archive for Space Telescopes (MAST) at the Space Telescope Science Institute. The specific observations analyzed can be accessed via \dataset[http://dx.doi.org/10.17909/agda-2w34]{http://dx.doi.org/10.17909/agda-2w34}. STScI is operated by the Association of Universities for Research in Astronomy, Inc., under NASA contract NAS5–26555. Support to MAST for these data is provided by the NASA Office of Space Science via grant NAG5–7584 and by other grants and contracts.
This research made use of \textsc{Photutils}, an Astropy package for
detection and photometry of astronomical sources \citep{bradley20}.

\software{
{\sc astropy} \fst{\citep{astropy13, astropy, astropy22}},
{\sc astrorms},
{\sc jwst pipeline}, 
{\sc photutils} \citep{bradley20},
{\sc pypher} \citep{boucaud16},
{\sc webbpsf} \fst{\citep{webbpsf15}}, 
\fst{ {\sc scipy} \citep{scipy20} },
{\sc source extractor} \citep{bertin96},
{\sc sep} \citep{barbary16},
{\sc tphot} \citep{merlin15, merlin16}
}

\bibliography{all}{}
\bibliographystyle{aasjournal}

\end{CJK*}
\end{document}